\input epsf
\expandafter\edef\csname hypers@fe\endcsname{\catcode
                                             `\noexpand @=\the\catcode`\@}%
\catcode`\@=11
%
%
\ifx\hyperd@ne\hyper@ndefined
 \global\let\hyperd@ne=\relax
\else
 \errhelp{hyperbasics.tex needs to be included only once outside
          of any {...} or \begingroup...\endgroup. You have tried to
          include it more than once. If the previous include was indeed
          outside any groupings, continue and all will be well.}%
  
\fi
%
%
\def\hyperv@rsion{8}%
%
%
\newread\hyperf@le
\def\hyperf@lename{\jobname.hrf}%
\immediate\openin\hyperf@le\hyperf@lename\relax
\ifeof\hyperf@le\relax
 \immediate\closein\hyperf@le\relax
\else
 \immediate\closein\hyperf@le\relax
 \input \hyperf@lename
\fi
%
%
\newwrite\hyperf@le
\immediate\openout\hyperf@le\hyperf@lename
%
%
\newtoks\hypert@ks
%
%
\edef\hypert@mp{\catcode`\noexpand\#=\the\catcode`\#}%
\catcode`\#=12
\def\hyperh@sh{#}%
\hypert@mp
\let\hypert@mp=\relax
\let\hyper@nd=\relax
\def\hyperstr@pquote"#1"#2\hyper@nd{\ifx\hyper@ndefined#2\hyper@ndefined#1\else
                                    \ifx\hyper@ndefined#1\hyper@ndefined
                                    \hyperstr@pquote#2"\hyper@nd\else
                                    #1\hyperstr@pquote"#2"\hyper@nd\fi\fi}%
\def\hyperstr@pblank" #1 #2\hyper@nd"{\ifx\hyper@ndefined#2\hyper@ndefined#1\else
                                    \ifx\hyper@ndefined#1\hyper@ndefined
                                    \hyperstr@pblank"#2 \hyper@nd"\else
                                    #1\hyperstr@pblank" #2 \hyper@nd"\fi\fi}
\long\def\hyper@nchor#1#2{\edef\hyperm@cro{html:<A #1>}%
                          \special\expandafter{\hyperm@cro}%
                          {#2}}%
\def\hyper@atm@ning#1->#2\hyper@nd{#2}
\def\hyperlink#1{\edef\hypert@mp{#1}%
               \edef\hypert@mp{\expandafter\hyper@atm@ning\meaning\hypert@mp
                               \hyper@nd}%
               \edef\hypert@mp"{ \expandafter\hyperstr@pquote\expandafter"%
                               \hypert@mp"\hyper@nd}%
               \edef\hypert@mp{\expandafter\hyperstr@pblank\expandafter%
                               "\hypert@mp" \hyper@nd"}%
               \hyper@nchor{href=\expandafter"\hypert@mp"}}%
\def\hypertarget#1{\edef\hypert@mp{#1}%
               \edef\hypert@mp{\expandafter\hyper@atm@ning\meaning\hypert@mp
                               \hyper@nd}%
               \edef\hypert@mp"{ \expandafter\hyperstr@pquote\expandafter"%
                               \hypert@mp"\hyper@nd}%
               \edef\hypert@mp{\expandafter\hyperstr@pblank\expandafter%
                               "\hypert@mp" \hyper@nd"}%
               \hyper@nchor{name=\expandafter"\hypert@mp"}}%
\def\hyperref{\afterassignment\hyperr@f\let\hyperp@ram}
\def\hyperr@f{\ifx\hyperp@ram{\iffalse}\fi
               \expandafter\expandafter\expandafter\hyperr@@
               \expandafter{%
              \else
               \iffalse}\fi
               \ifx\hyperp@ram\hyper@ndefined
                 \message{Undefined reference}%
                 \def\hyperp@r@m{{}{undefined}{}}%
               \else
                 \edef\hyperp@r@m{\hyperp@ram}%
               \fi
               \expandafter\expandafter\expandafter\hyperr@@
               \expandafter\hyperp@r@m
              \fi}%
\def\hyperr@@#1#2#3{\ifx\hyper@ndefined#1\hyper@ndefined
                    \hypert@ks\expandafter{\hyperh@sh#2.#3}%
                    \else
                     \ifx\hyper@ndefined#2#3\hyper@ndefined
                      \hypert@ks{#1}%
                     \else
                      \def\hypert@mp{#1}%
                      \hypert@ks\expandafter\expandafter\expandafter
                      {\expandafter\hypert@mp\hyperh@sh#2.#3}%
                     \fi
                    \fi
                    \expandafter\hyperlink\expandafter{\the\hypert@ks}}%
\def\hyperdef#1#2#3{{\global\escapechar=`\\\relax
                     \edef\hypert@mp{\hyperstr@pquote"#2.#3"\hyper@nd}%
                     \expandafter\ifx\csname hyperd@\meaning\hypert@mp
                     \endcsname
                     \relax
                     \expandafter\gdef\csname hyperd@\meaning\hypert@mp
                     \endcsname{}%
                     \gdef#1{{}{\hyperstr@pquote"#2"\hyper@nd}%
                               {\hyperstr@pquote"#3"\hyper@nd}}%
                     \immediate\write\hyperf@le{\def\noexpand#1{#1}}%
                     \xdef\hypert@mp{\global\let\noexpand\hypert@mp=\relax
                                     \noexpand\hypertarget{\hypert@mp}}%
                     \global\hypert@ks={\hypert@mp}%
                     \else
                     \message\expandafter{'\hypert@mp' duplicate}%
                     \global\let\hypert@mp=\relax
                     \global\hypert@ks={\hyperdef{#1}{#2}{#3@}}%
                     \fi}\the\hypert@ks}%

\def\hyper@nique#1#2#3#4{\global\escapechar=`\\\relax
                     \edef\hypert@mp{\hyperstr@pquote"#2.#3"\hyper@nd}%
                     \expandafter\ifx\csname hyperd@\meaning\hypert@mp
                     \endcsname
                     \relax
                     \gdef#1{{}{\hyperstr@pquote"#2"\hyper@nd}%
                               {\hyperstr@pquote"#3"\hyper@nd}}%
                     \global\let\hypert@mp=\relax
                     #4%
                     \else
                     \global\let\hypert@mp=\relax
                     \hyper@nique{#1}{#2}{#3@}{#4}%
                     \fi
                     }%

\let\hyper@@@@=\relax
\def\hyper@@{\let\hyper@@@=\relax}%
\hyper@@
\def\hyper@{\relax\let\hyper@@@\noexpand\hyper@\noexpand}%
\def\hyperpr@ref{\hyper@@\hyperref}
\def\hyperpr@def{\hyper@@\hyperdef}

\let\href\hyperlink

%
%
\hypers@fe


\def\sla#1{\mkern-1.5mu\raise0.4pt\hbox{$\not$}\mkern1.2mu #1\mkern 0.7mu}
\def\Dbar{\mkern-1.5mu\raise0.4pt\hbox{$\not$}\mkern-.1mu {\rm D}\mkern.1mu}
\def\Abar{\mkern1.mu\raise0.4pt\hbox{$\not$}\mkern-1.3mu A\mkern.1mu}
\def\Bbar{\mkern-0.mu\raise0.4pt\hbox{$\not$}\mkern-.3mu B\mkern 0.6mu}
\newskip\tableskipamount \tableskipamount=8pt plus 3pt minus 3pt

\newdimen\chapskip
\chapskip=17.5mm
\font\twbfx=cmbx10 scaled 1200
\font\chapfnt=cmbx10 scaled 1440
\font\chapbxten=cmbx10
\font\chapbxseven=cmbx7

\font\ssbx=cmssbx10  

\font\chaprm=cmr10 scaled 1440
\font\chaprmscript=cmr10
\font\chaprmseven=cmr7
\font\chapssfnt=cmssbx10 scaled 1440

\font\chapibfnt=cmmib10 scaled 1440
\font\chapmifnt=cmmi10 scaled 1440
\font\chapsyfnt=cmsy10 scaled 1440
\font\chapexfnt=cmex10 scaled 1440
\font\chapibten=cmmib10
\font\chapibseven=cmmib7
\font\chapmiscript=cmmi10  
\font\chapsyscript=cmsy10  
\font\chapexscript=cmex10
\font\chapmiseven=cmmi7
\font\chapsyseven=cmsy7
\font\chapexseven=cmex7  
\def\chapfont{
\textfont0=\chaprm\scriptfont0=\chaprmscript\scriptscriptfont0=\chaprmseven
\textfont1=\chapmifnt \scriptfont1=\chapmiscript  \scriptscriptfont1=\chapmiseven
\textfont2=\chapsyfnt \scriptfont2=\chapsyscript\scriptscriptfont2=\chapsyseven
\textfont3=\chapexfnt \scriptfont3=\chapexscript \scriptscriptfont3=\chapexseven
\textfont\bffam=\chapfnt \scriptfont\bffam=\chapbxten
\scriptscriptfont\bffam=\chapbxseven
\def\bf{\fam\bffam\chapfnt} 
\textfont4=\chapibfnt \scriptfont4=\chapibten \scriptscriptfont4=\chapibseven
\abovedisplayskip=17pt plus 5pt minus 13pt
\belowdisplayskip=\abovedisplayskip
\normalbaselineskip=17pt
\setbox\strutbox=\hbox{\vrule height12.2pt depth5.0pt width0pt}
\normalbaselines \chapssfnt}

\font\caprm=cmr9
\font\capit=cmti9
\font\capbf=cmbx9
\font\capsl=cmsl9
\font\capmi=cmmi9
\font\capex=cmex9
\font\capsy=cmsy9

\def\makeheadline{\vbox to 0pt{\vskip-22.5pt
\line{\vbox to8.5pt{}\the\headline}\vss}\nointerlineskip}
\font\headrm=cmr10

\font\sixrm=cmr6
\font\fiverm=cmr5
\font\sixmi=cmmi6
\font\fivemi=cmmi5
\font\sixsy=cmsy6
\font\fivesy=cmsy5
\font\sixbf=cmbx6
\font\fivebf=cmbx5
\skewchar\capmi='177 \skewchar\sixmi='177 \skewchar\fivemi='177
\skewchar\capsy='60 \skewchar\sixsy='60 \skewchar\fivesy='60


%
%

\catcode`\@=11
\font\tenbi=cmmib10 
\font\ninebi=cmmib9
\font\sevenbi=cmmib7 
\font\fivebi=cmmib5
\textfont4=\tenbi \scriptfont4=\sevenbi \scriptscriptfont4=\fivebi
\newfam\mibfam
\textfont\mibfam=\tenbi \scriptfont\mibfam=\sevenbi \scriptscriptfont\mibfam=\fivebi
\def\Bfg{\ifmmode\let\next\Bfg@\else
 \def\next{\errmessage{Use \string\Bfg\space only in math mode}}\fi\next}
\def\Bfg@#1{{\Bfg@@{#1}}}
\def\Bfg@@#1{\fam\mibfam#1}
\skewchar\tenbi='177\skewchar\sevenbi='177
\mathchardef\alpha="710B
\mathchardef\beta="710C 
\mathchardef\gamma="710D
\mathchardef\delta="710E
\mathchardef\epsilon="710F  
\mathchardef\zeta="7110  
\mathchardef\eta="7111 
\mathchardef\theta="7112 
\mathchardef\iota="7113 
\mathchardef\kappa="7114 
\mathchardef\lambda="7115 
\mathchardef\mu="7116 
\mathchardef\nu="7117 
\mathchardef\xi="7118 
\mathchardef\pi="7119 
\mathchardef\rho="711A  
\mathchardef\sigma="711B 
\mathchardef\tau="711C 
\mathchardef\upsilon="711D 
\mathchardef\phi="711E 
\mathchardef\chi="711F 
\mathchardef\psi="7120 
\mathchardef\omega="7121 
\mathchardef\varepsilon="7122 
\mathchardef\vartheta="7123
\mathchardef\varpi="7124 
\mathchardef\varrho="7125 
\mathchardef\varsigma="7126 
\mathchardef\varphi="7127
\mathchardef\alphab="040B
\mathchardef\betab="040C 
\mathchardef\gammab="040D
\mathchardef\deltab="040E
\mathchardef\epsilonb="040F  
\mathchardef\zetab="0410  
\mathchardef\etab="0411 
\mathchardef\thetab="0412 
\mathchardef\iotab="0413 
\mathchardef\kappab="0414 
\mathchardef\lambdab="0415 
\mathchardef\mub="0416 
\mathchardef\nub="0417 
\mathchardef\xib="0418 
\mathchardef\pib="0419 
\mathchardef\rhob="041A  
\mathchardef\sigmab="041B 
\mathchardef\taub="041C 
\mathchardef\upsilonb="041D 
\mathchardef\phib="041E 
\mathchardef\chib="041F 
\mathchardef\psib="0420 
\mathchardef\omegab="0421 
\mathchardef\varepsilonb="0422 
\mathchardef\varthetab="0423
\mathchardef\varpib="0424 
\mathchardef\varrhob="0425 
\mathchardef\varsigmab="0426 
\mathchardef\varphib="0427 
\font\tenmsa=msam10
\font\sevenmsa=msam7
\font\fivemsa=msam5
\font\tenmsb=msbm10
\font\sevenmsb=msbm7
\font\fivemsb=msbm5
\newfam\msafam
\newfam\msbfam
\textfont\msafam=\tenmsa  \scriptfont\msafam=\sevenmsa
  \scriptscriptfont\msafam=\fivemsa
\textfont\msbfam=\tenmsb  \scriptfont\msbfam=\sevenmsb
  \scriptscriptfont\msbfam=\fivemsb

\def\hexnumber@#1{\ifcase#1 0\or1\or2\or3\or4\or5\or6\or7\or8\or9\or
	A\or B\or C\or D\or E\or F\fi }
\font\teneuf=eufm10
\font\seveneuf=eufm7
\font\fiveeuf=eufm5
\newfam\euffam
\textfont\euffam=\teneuf
\scriptfont\euffam=\seveneuf
\scriptscriptfont\euffam=\fiveeuf
\def\frak{\ifmmode\let\next\frak@\else
 \def\next{\Err@{Use \string\frak\space only in math mode}}\fi\next}
\def\goth{\ifmmode\let\next\frak@\else
 \def\next{\Err@{Use \string\goth\space only in math mode}}\fi\next}
\def\frak@#1{{\frak@@{#1}}}
\def\frak@@#1{\fam\euffam#1}

\edef\msa@{\hexnumber@\msafam}
\edef\msb@{\hexnumber@\msbfam}

\mathchardef\boxdot="2\msa@00
\mathchardef\boxplus="2\msa@01
\mathchardef\boxtimes="2\msa@02
\mathchardef\square="0\msa@03
\mathchardef\blacksquare="0\msa@04
\mathchardef\centerdot="2\msa@05
\mathchardef\lozenge="0\msa@06
\mathchardef\blacklozenge="0\msa@07
\mathchardef\circlearrowright="3\msa@08
\mathchardef\circlearrowleft="3\msa@09
\mathchardef\rightleftharpoons="3\msa@0A
\mathchardef\leftrightharpoons="3\msa@0B
\mathchardef\boxminus="2\msa@0C
\mathchardef\Vdash="3\msa@0D
\mathchardef\Vvdash="3\msa@0E
\mathchardef\vDash="3\msa@0F
\mathchardef\twoheadrightarrow="3\msa@10
\mathchardef\twoheadleftarrow="3\msa@11
\mathchardef\leftleftarrows="3\msa@12
\mathchardef\rightrightarrows="3\msa@13
\mathchardef\upuparrows="3\msa@14
\mathchardef\downdownarrows="3\msa@15
\mathchardef\upharpoonright="3\msa@16

\mathchardef\downharpoonright="3\msa@17
\mathchardef\upharpoonleft="3\msa@18
\mathchardef\downharpoonleft="3\msa@19
\mathchardef\rightarrowtail="3\msa@1A
\mathchardef\leftarrowtail="3\msa@1B
\mathchardef\leftrightarrows="3\msa@1C
\mathchardef\rightleftarrows="3\msa@1D
\mathchardef\Lsh="3\msa@1E
\mathchardef\Rsh="3\msa@1F
\mathchardef\rightsquigarrow="3\msa@20
\mathchardef\leftrightsquigarrow="3\msa@21
\mathchardef\looparrowleft="3\msa@22
\mathchardef\looparrowright="3\msa@23
\mathchardef\circeq="3\msa@24
\mathchardef\succsim="3\msa@25
\mathchardef\gtrsim="3\msa@26
\mathchardef\gtrapprox="3\msa@27
\mathchardef\multimap="3\msa@28
\mathchardef\therefore="3\msa@29
\mathchardef\because="3\msa@2A
\mathchardef\doteqdot="3\msa@2B

\mathchardef\triangleq="3\msa@2C
\mathchardef\precsim="3\msa@2D
\mathchardef\lesssim="3\msa@2E
\mathchardef\lessapprox="3\msa@2F
\mathchardef\eqslantless="3\msa@30
\mathchardef\eqslantgtr="3\msa@31
\mathchardef\curlyeqprec="3\msa@32
\mathchardef\curlyeqsucc="3\msa@33
\mathchardef\preccurlyeq="3\msa@34
\mathchardef\leqq="3\msa@35
\mathchardef\leqslant="3\msa@36
\mathchardef\lessgtr="3\msa@37
\mathchardef\backprime="0\msa@38
\mathchardef\risingdotseq="3\msa@3A
\mathchardef\fallingdotseq="3\msa@3B
\mathchardef\succcurlyeq="3\msa@3C
\mathchardef\geqq="3\msa@3D
\mathchardef\geqslant="3\msa@3E
\mathchardef\gtrless="3\msa@3F
\mathchardef\sqsubset="3\msa@40
\mathchardef\sqsupset="3\msa@41
\mathchardef\vartriangleright="3\msa@42
\mathchardef\vartriangleleft="3\msa@43
\mathchardef\trianglerighteq="3\msa@44
\mathchardef\trianglelefteq="3\msa@45
\mathchardef\bigstar="0\msa@46
\mathchardef\between="3\msa@47
\mathchardef\blacktriangledown="0\msa@48
\mathchardef\blacktriangleright="3\msa@49
\mathchardef\blacktriangleleft="3\msa@4A
\mathchardef\vartriangle="0\msa@4D
\mathchardef\blacktriangle="0\msa@4E
\mathchardef\triangledown="0\msa@4F
\mathchardef\eqcirc="3\msa@50
\mathchardef\lesseqgtr="3\msa@51
\mathchardef\gtreqless="3\msa@52
\mathchardef\lesseqqgtr="3\msa@53
\mathchardef\gtreqqless="3\msa@54
\mathchardef\Rrightarrow="3\msa@56
\mathchardef\Lleftarrow="3\msa@57
\mathchardef\veebar="2\msa@59
\mathchardef\barwedge="2\msa@5A
\mathchardef\doublebarwedge="2\msa@5B
\mathchardef\angle="0\msa@5C
\mathchardef\measuredangle="0\msa@5D
\mathchardef\sphericalangle="0\msa@5E
\mathchardef\varpropto="3\msa@5F
\mathchardef\smallsmile="3\msa@60
\mathchardef\smallfrown="3\msa@61
\mathchardef\Subset="3\msa@62
\mathchardef\Supset="3\msa@63
\mathchardef\Cup="2\msa@64

\mathchardef\Cap="2\msa@65

\mathchardef\curlywedge="2\msa@66
\mathchardef\curlyvee="2\msa@67
\mathchardef\leftthreetimes="2\msa@68
\mathchardef\rightthreetimes="2\msa@69
\mathchardef\subseteqq="3\msa@6A
\mathchardef\supseteqq="3\msa@6B
\mathchardef\bumpeq="3\msa@6C
\mathchardef\Bumpeq="3\msa@6D
\mathchardef\lll="3\msa@6E

\mathchardef\ggg="3\msa@6F

\mathchardef\circledS="0\msa@73
\mathchardef\pitchfork="3\msa@74
\mathchardef\dotplus="2\msa@75
\mathchardef\backsim="3\msa@76
\mathchardef\backsimeq="3\msa@77
\mathchardef\complement="0\msa@7B
\mathchardef\intercal="2\msa@7C
\mathchardef\circledcirc="2\msa@7D
\mathchardef\circledast="2\msa@7E
\mathchardef\circleddash="2\msa@7F
\def\ulcorner{\delimiter"4\msa@70\msa@70 }
\def\urcorner{\delimiter"5\msa@71\msa@71 }
\def\llcorner{\delimiter"4\msa@78\msa@78 }
\def\lrcorner{\delimiter"5\msa@79\msa@79 }
\def\yen{\mathhexbox\msa@55 }
\def\checkmark{\mathhexbox\msa@58 }
\def\circledR{\mathhexbox\msa@72 }
\def\maltese{\mathhexbox\msa@7A }
\mathchardef\lvertneqq="3\msb@00
\mathchardef\gvertneqq="3\msb@01
\mathchardef\nleq="3\msb@02
\mathchardef\ngeq="3\msb@03
\mathchardef\nless="3\msb@04
\mathchardef\ngtr="3\msb@05
\mathchardef\nprec="3\msb@06
\mathchardef\nsucc="3\msb@07
\mathchardef\lneqq="3\msb@08
\mathchardef\gneqq="3\msb@09
\mathchardef\nleqslant="3\msb@0A
\mathchardef\ngeqslant="3\msb@0B
\mathchardef\lneq="3\msb@0C
\mathchardef\gneq="3\msb@0D
\mathchardef\npreceq="3\msb@0E
\mathchardef\nsucceq="3\msb@0F
\mathchardef\precnsim="3\msb@10
\mathchardef\succnsim="3\msb@11
\mathchardef\lnsim="3\msb@12
\mathchardef\gnsim="3\msb@13
\mathchardef\nleqq="3\msb@14
\mathchardef\ngeqq="3\msb@15
\mathchardef\precneqq="3\msb@16
\mathchardef\succneqq="3\msb@17
\mathchardef\precnapprox="3\msb@18
\mathchardef\succnapprox="3\msb@19
\mathchardef\lnapprox="3\msb@1A
\mathchardef\gnapprox="3\msb@1B
\mathchardef\nsim="3\msb@1C
\mathchardef\ncong="3\msb@1D

\mathchardef\varsubsetneq="3\msb@20
\mathchardef\varsupsetneq="3\msb@21
\mathchardef\nsubseteqq="3\msb@22
\mathchardef\nsupseteqq="3\msb@23
\mathchardef\subsetneqq="3\msb@24
\mathchardef\supsetneqq="3\msb@25
\mathchardef\varsubsetneqq="3\msb@26
\mathchardef\varsupsetneqq="3\msb@27
\mathchardef\subsetneq="3\msb@28
\mathchardef\supsetneq="3\msb@29
\mathchardef\nsubseteq="3\msb@2A
\mathchardef\nsupseteq="3\msb@2B
\mathchardef\nparallel="3\msb@2C
\mathchardef\nmid="3\msb@2D
\mathchardef\nshortmid="3\msb@2E
\mathchardef\nshortparallel="3\msb@2F
\mathchardef\nvdash="3\msb@30
\mathchardef\nVdash="3\msb@31
\mathchardef\nvDash="3\msb@32
\mathchardef\nVDash="3\msb@33
\mathchardef\ntrianglerighteq="3\msb@34
\mathchardef\ntrianglelefteq="3\msb@35
\mathchardef\ntriangleleft="3\msb@36
\mathchardef\ntriangleright="3\msb@37
\mathchardef\nleftarrow="3\msb@38
\mathchardef\nrightarrow="3\msb@39
\mathchardef\nLeftarrow="3\msb@3A
\mathchardef\nRightarrow="3\msb@3B
\mathchardef\nLeftrightarrow="3\msb@3C
\mathchardef\nleftrightarrow="3\msb@3D
\mathchardef\divideontimes="2\msb@3E
\mathchardef\varnothing="0\msb@3F
\mathchardef\nexists="0\msb@40
\mathchardef\mho="0\msb@66
\mathchardef\eth="0\msb@67
\mathchardef\eqsim="3\msb@68
\mathchardef\beth="0\msb@69
\mathchardef\gimel="0\msb@6A
\mathchardef\daleth="0\msb@6B
\mathchardef\lessdot="3\msb@6C
\mathchardef\gtrdot="3\msb@6D
\mathchardef\ltimes="2\msb@6E
\mathchardef\rtimes="2\msb@6F
\mathchardef\shortmid="3\msb@70
\mathchardef\shortparallel="3\msb@71
\mathchardef\smallsetminus="2\msb@72
\mathchardef\thicksim="3\msb@73
\mathchardef\thickapprox="3\msb@74
\mathchardef\approxeq="3\msb@75
\mathchardef\succapprox="3\msb@76
\mathchardef\precapprox="3\msb@77
\mathchardef\curvearrowleft="3\msb@78
\mathchardef\curvearrowright="3\msb@79
\mathchardef\digamma="0\msb@7A
\mathchardef\varkappa="0\msb@7B
\mathchardef\hslash="0\msb@7D
\mathchardef\hbar="0\msb@7E
\mathchardef\backepsilon="3\msb@7F
\def\Bbb{\ifmmode\let\next\Bbb@\else
 \def\next{\errmessage{Use \string\Bbb\space only in math mode}}\fi\next}
\def\Bbb@#1{{\Bbb@@{#1}}}
\def\Bbb@@#1{\fam\msbfam#1}

\def\elevenpoint{
\textfont0=\caprm \scriptfont0=\sixrm \scriptscriptfont0=\fiverm
\def\rm{\fam0\caprm}
\textfont1=\capmi \scriptfont1=\sixmi \scriptscriptfont1=\fivemi
\textfont2=\capsy \scriptfont2=\sixsy \scriptscriptfont2=\fivesy
\textfont3=\capex \scriptfont3=\capex \scriptscriptfont3=\capex
\textfont\itfam=\capit \def\it{\fam\itfam\capit} 
\textfont\slfam=\capsl  \def\sl{\fam\slfam\capsl} 
\textfont\bffam=\capbf \scriptfont\bffam=\sixbf
\scriptscriptfont\bffam=\fivebf
\def\bf{\fam\bffam\capbf} 
\textfont4=\ninebi \scriptfont4=\sevenbi \scriptscriptfont4=\fivebi
\abovedisplayskip=11pt plus 3pt minus 8pt
\belowdisplayskip=\abovedisplayskip
\smallskipamount=2.7pt plus 1pt minus 1pt
\medskipamount=5.4pt plus 2pt minus 2pt
\bigskipamount=10.8pt plus 3.6pt minus 3.6pt
\normalbaselineskip=11pt
\setbox\strutbox=\hbox{\vrule height7.8pt depth3.2pt width0pt}
\normalbaselines \rm}
\catcode`\@=12


\catcode`\@=11
\def\unredoffs{\voffset=13mm \hoffset=6.5truemm}
\def\redoffs{\voffset=-12.truemm\hoffset=-3truemm}

\newif\ifbookmode
\bookmodefalse
%
\newwrite\inx
\def\book{\bookmodetrue\immediate\openout\inx=\jobname.inx%
 \hsize=120mm\vsize=195mm}
\def\@#1{\noindent\ifbookmode\write\inx{#1,\space
\number\pageno.\par}\fi}

\newbox\leftpage \newdimen\fullhsize \newdimen\hstitle \newdimen\hsbody
\newdimen\hdim
\hfuzz=1pt
\ifx\hyperdef\UNd@FiNeD\def\hyperdef#1#2#3#4{#4}\def\hyperref#1#2#3#4{#4}\fi
\def\bigans{b }
\def\answ{b }
\ifx\answ\bigans\message{(Format simple colonne 12pts.}
\magnification=1200 \unredoffs
\hsize=122.5mm\vsize=182.5mm
\hsbody=\hsize \hstitle=\hsize 
\else\message{(Format simple colonne, 10pts.} \let\l@r=L
\magnification=1000
\redoffs%
\hsize=122.5mm\vsize=182.5mm
\hsbody=\hsize \hstitle=\hsize 
\fi
%


%
\def\sla#1{\mkern-1.5mu\raise0.4pt\hbox{$\not$}\mkern1.2mu #1\mkern 0.7mu}
\def\Dbar{\mkern-1.5mu\raise0.4pt\hbox{$\not$}\mkern-.1mu {\rm D}\mkern.1mu}
\def\Abar{\mkern1.mu\raise0.4pt\hbox{$\not$}\mkern-1.3mu A\mkern.1mu}
\nopagenumbers
\def\makeheadline{\vbox to 0pt{\vskip-27pt
\line{\vbox to8.5pt{}\the\headline}\vss}\nointerlineskip}
\def\subsectionname{}
\def\sectionname{}
\headline={\ifnum\pageno=1\hfill\else\draftdate\hfil{\headrm\folio}%
\hfil\hphantom{\draftdate}\fi}


\newcount\yearltd\yearltd=\year\advance\yearltd by -2000
\newif\ifdraftmode
\draftmodefalse
\def\draft{\draftmodetrue{\count255=\time\divide\count255 by 60
\xdef\hourmin{\number\count255}
  \multiply\count255 by-60\advance\count255 by\time
  \xdef\hourmin{\hourmin:\ifnum\count255<10 0\fi\the\count255}}}
\def\draftdate{\ifdraftmode{\headrm\quad (\jobname,\ le
\number\day/\number\month/\number\yearltd\ \ \hourmin)}\else{}\fi}
\newif\iffrancmode
\francmodefalse

\def\sgn{\mathop{\rm sgn}\nolimits}

\def\d{{\rm d}}
\def\ud{{\textstyle{1\over 2}}}
\def\half{\ud}

\def\del{\partial}

\chardef\sigmat=27

\def\frac#1#2{{\textstyle{#1\over#2}}}

\def\leaderfill{\leaders\hbox to 1em{\hss.\hss}\hfill}
\catcode`\@=11

\def\deqalignno#1{\displ@y\tabskip\centering \halign to
\displaywidth{\hfil$\displaystyle{##}$\tabskip0pt&$\displaystyle{{}##}$
\hfil\tabskip0pt &\quad
\hfil$\displaystyle{##}$\tabskip0pt&$\displaystyle{{}##}$
\hfil\tabskip\centering& \llap{$##$}\tabskip0pt \crcr #1 \crcr}}
\def\deqalign#1{\null\,\vcenter{\openup\jot\m@th\ialign{
\strut\hfil$\displaystyle{##}$&$\displaystyle{{}##}$\hfil
&&\quad\strut\hfil$\displaystyle{##}$&$\displaystyle{{}##}$
\hfil\crcr#1\crcr}}\,}
\def\xlabel#1{\expandafter\xl@bel#1}\def\xl@bel#1{#1}
\def\label#1{\l@bel #1{\hbox{}}}
\def\l@bel#1{\ifx\UNd@FiNeD#1\message{label \string#1 is undefined.}%
\xdef#1{?.? }\fi{\let\hyperref=\relax\xdef\next{#1}}%
\ifx\next\em@rk\def\next{}%
\else\def\next{#1}\fi\next}
\def\DefWarn#1{\ifx\UNd@FiNeD#1\else
\immediate\write16{*** WARNING: the label \string#1 is already defined%
***}\fi}%
\newread\ch@ckfile
\def\cinput#1{\def\filen@me{#1 }
\immediate\openin\ch@ckfile=\filen@me
\ifeof\ch@ckfile\message{<< (\filen@me\ DOES NOT EXIST in this pass)>>}\else%
\closein \ch@ckfile\input\filen@me\fi}
\ifx\UNd@FiNeD\prefix\def\prefix{}\fi 
\newread\ch@ckfile
\immediate\openin\ch@ckfile=\jobname.def
\ifeof\ch@ckfile\message{<< (\jobname.def DOES NOT EXIST in this pass) >>}
\else
\def\DefWarn#1{}%
\closein \ch@ckfile%
\input\jobname.def\fi
\def\listcontent{
\immediate\openin\ch@ckfile=\jobname.tab 
\ifeof\ch@ckfile\message{no file \jobname.tab, no table of contents this
pass}%
\else\closein\ch@ckfile%
\def\sectionname{\iffrancmode Table des
mati\`eres \else Contents\fi}
\centerline{\twbfx\iffrancmode Table des
mati\`eres \else Contents\fi}\nobreak\medskip%
{\baselineskip=12pt\parskip=0pt\catcode`\@=11\input\jobname.tab
\catcode`\@=12\bigbreak\bigskip}\fi}
\newcount\nochapter
\newcount\nosection
\newcount\nosubsection
\newcount\neqno
\newcount\notenumber
\newcount\nofigure
\newcount\notable
\newcount\noexerc
\newcount\fpage
\newcount\firstpage
\newif\ifappmode
\newwrite\equa

\newdimen\hulp
\def\maketitle#1{
\edef\oneliner##1{\centerline{##1}}
\edef\twoliner##1{\vbox{\parindent=0pt\leftskip=0pt plus 1fill\rightskip=0pt
plus 1fill
                     \parfillskip=0pt\relax##1}}
\setbox0=\vbox{#1}\hulp=0.5\hsize
                 \ifdim\wd0<\hulp\oneliner{#1}\else
                 \twoliner{#1}\fi}
\def\preprint#1{\ifdraftmode\gdef\prepname{\jobname/#1}\else%
\gdef\prepname{#1}\fi\hfill{
\expandafter{\prepname}}\vskip20mm}
\def\title#1\par{\gdef\titlename{#1}
\global\firstpage=\pageno
\nosection=0
\mark{\the\nosection}
\maketitle{\ssbx\uppercase\expandafter{\titlename}}
\vskip17mm
\nochapter=0
\neqno=0
\notenumber=0
\nofigure=0
\notable=0
\def\prefix{}
\appmodefalse
\mark{\the\nochapter}
\message{#1}
\immediate\openout\equa=\jobname.equ %
}
\ifbookmode%
\headline={\ifnum\pageno=\firstpage\hfill\else\ifodd\pageno\rightheadline
\else\leftheadline\fi\fi}
\else
\headline={\ifnum\pageno=\firstpage\hfill\else\draftdate\hfil{\headrm\folio}%
\hfil\hphantom{\draftdate}\fi}\fi
\def\abstract{\vskip8mm\iffrancmode\centerline{R\'ESUM\'E}\else%
\centerline{ABSTRACT}\fi \smallskip \begingroup\narrower
\elevenpoint\baselineskip10pt}
\def\endabstract{\par\endgroup \bigskip}
\def\section#1\par{\vskip0pt plus.1\vsize\penalty-100\vskip0pt plus-.1
\vsize\bigskip\vglue\parskip\par%
\global\fpage=\pageno
\ifnum\nochapter=0\ifappmode\relax\else\writetoc
\fi\fi
\advance\nochapter by 1\global\nosection=0\global\neqno=0
\gdef\sectionname{#1}
\vbox{\noindent\twbfx{\hyperdef\hypernoname{section}{\prefix\the\nochapter}%
{\prefix\the\nochapter}\ #1}}%
\message{\prefix\the\nochapter\ \sectionname}%
\writetoca{{\string\hyperref{}{section}{\prefix\the\nochapter}%
{\prefix\the\nochapter}} {#1}}%
\bigskip\noindent%
}

\def\appendix#1#2\par{\bigbreak\nochapter=0
\appmodetrue
\notenumber=0
\neqno=0
\def\prefix{A}
\mark{\the\nochapter}
\message{APPENDICES}
{\hyperdef\hypernoname{section}{\prefix}{
\leftline{\uppercase\expandafter{#1}}
\leftline{\uppercase\expandafter{#2}}}}
\noindent\nonfrenchspacing
\writetoca{\string\hyperref{}{section}{\prefix}{Appendices}.\ #1 \ #2}%
}
\def\subsection#1\par{\vskip0pt plus.05\vsize\penalty-100\vskip0pt
plus-.05\vsize\bigskip\vskip1mm\advance\nosection by 1
\global\nosubsection=0
\def\subsectionname{#1}
\mark{#1}
\message{\the\nochapter.\the\nosection\ #1}
\vbox{\noindent\bf{\hyperdef\hypernoname{section}{\prefix\the\nochapter.%
\the\nosection}{\prefix\the\nochapter.\the\nosection\ #1}}}\vskip1mm%
\smallskip\noindent%
\writetoca{{\string\hyperref{}{section}{\prefix\the\nochapter.%
\the\nosection}{\prefix\the\nochapter.\the\nosection}} {#1}}%
}
\def\ssubsection#1\par{\vskip0pt plus.05\vsize\penalty-100\vskip0pt%
plus-.05\vsize\medskip\vskip\parskip\advance\nosubsection by 1%
\message{\the\nochapter.\the\nosection.\the\nosubsection\ #1}%
\vbox{\noindent\bf{\hyperdef\hypernoname{section}{\prefix\the\nochapter.%
\the\nosection.\the\nosubsection}{\prefix\the\nochapter.\the\nosection.%
\the\nosubsection\ \bf #1}}}%
\smallskip\noindent%
\writetoca{{\string\hyperref{}{section}{\prefix\the\nochapter.%
\the\nosection.\the\nosubsection}{\prefix\the\nochapter.\the\nosection.%
\the\nosubsection}} {#1}}%
}
%
\def\note #1{\advance\notenumber by 1
\footnote{$^{\the\notenumber}$}{\sevenrm #1}}

\parindent=1em
\newinsert\margin
\dimen\margin=\maxdimen
\count\margin=0 \skip\margin=0pt
\def\sslbl#1{\DefWarn#1%
\ifdraftmode{\leavevmode\vadjust{\smash%
{\line{{\escapechar=` \hfill\rlap{\sevenrm\hskip1mm\string#1}}}}}}%
\fi
\ifnum\nochapter=0%
\if\prefix{}\xdef#1{}%
\edef\ewrite{\write\equa{{\string#1}}%
\write\equa{}}\ewrite%
\else
\xdef#1{\noexpand\hyperref{}{section}{\prefix}{\prefix}}%
\edef\ewrite{\write\equa{{\string#1},\prefix}%
\write\equa{}}\ewrite%
\writedef{#1\leftbracket#1}
\fi
\else%
\ifnum\nosection=0%
\xdef#1{\noexpand\hyperref{}{section}{\prefix\the\nochapter}%
{\prefix\the\nochapter}}%
\edef\ewrite{\write\equa{{\string#1},\prefix\the\nochapter}%
\write\equa{}}\ewrite%
\writedef{#1\leftbracket#1}
\else%
\ifnum\nosubsection=0%
\xdef#1{\noexpand\hyperref{}{section}{\prefix\the\nochapter.%
\the\nosection}{\prefix\the\nochapter.\the\nosection}}%
\writedef{#1\leftbracket#1}
\edef\ewrite{\write\equa{{\string#1},\prefix\the\nochapter%
.\the\nosection}\write\equa{}}\ewrite%
\else%
\xdef#1{\noexpand\hyperref{}{section}%
{\prefix\the\nochapter.\the\nosection.\the\nosubsection}%
{\prefix\the\nochapter.\the\nosection.\the\nosubsection}}%
\writedef{#1\leftbracket#1}
\edef\ewrite{\write\equa{{\string#1},\prefix\the\nochapter.\the\nosection%
.\the\nosubsection}\write\equa{}}\ewrite%
\fi\fi\fi}%

\newwrite\tfile \def\writetoca#1{}
\def\writetoc{\immediate\openout\tfile=\jobname.tab
\def\writetoca##1{{\edef\next{\write\tfile{\noindent ##1 \string\leaderfill%
\noexpand\number\pageno\par}}\next}}}

%
\def\nolabels{\def\wrlabeL##1{}\def\eqlabeL##1{}\def\reflabeL##1{}}
\def\writelabels{\def\wrlabeL##1{\leavevmode\vadjust{\rlap{\smash%
{\line{{\escapechar=` \hfill\rlap{\sevenrm\hskip.03in\string##1}}}}}}}%
\def\eqlabeL##1{{\escapechar-1\rlap{\sevenrm\hskip.05in\string##1}}}%
\def\reflabeL##1{\noexpand\llap{\noexpand\sevenrm\string\string\string##1}}}
\ifdraftmode\writelabels\else\nolabels\fi

\global\newcount\refno \global\refno=1
\newwrite\rfile
\def\ref{[\hyperref{}{reference}{\the\refno}{\the\refno}]\nref}
\def\nref#1{\DefWarn#1%
\xdef#1{[\noexpand\hyperref{}{reference}{\the\refno}{\the\refno}]}%
\writedef{#1\leftbracket#1}%
\ifnum\refno=1\immediate\openout\rfile=\jobname.ref\fi
\chardef\wfile=\rfile\immediate\write\rfile{\noexpand\item{[\noexpand\hyperdef%
\noexpand\hypernoname{reference}{\the\refno}{\the\refno}]\ }%
\reflabeL{#1\hskip.31in}\pctsign}\global\advance\refno by1\findarg}
\def\findarg#1#{\begingroup\obeylines\newlinechar=`\^^M\pass@rg}
{\obeylines\gdef\pass@rg#1{\writ@line\relax #1^^M\hbox{}^^M}%
\gdef\writ@line#1^^M{\expandafter\toks0\expandafter{\striprel@x #1}%
\edef\next{\the\toks0}\ifx\next\em@rk\let\next=\endgroup\else\ifx\next\empty%
\else\immediate\write\wfile{\the\toks0}\fi\let\next=\writ@line\fi\next\relax}}
\def\striprel@x#1{} \def\em@rk{\hbox{}}
\def\lref{\begingroup\obeylines\lr@f}
\def\lr@f#1#2{\DefWarn#1\gdef#1{\let#1=\UNd@FiNeD\ref#1{#2}}\endgroup\unskip}

\def\addref#1{\immediate\write\rfile{\noexpand\item{}#1}} 
\def\listrefs{{}\vfill\supereject\immediate\closeout\rfile\writestoppt
\baselineskip=14pt
\gdef\reference{\iffrancmode  R\'eferences \else References\fi}
\mark{\reference}
\nochapter=0\def\sectionname{\reference}
{\centerline{\bf \hyperdef\hypernoname{refer}{refer}{\bf \reference}}}
\nobreak\bigskip\noindent
\writetoca{\string\hyperref{}{refer}{refer}{\reference}}
{\parindent=20pt%
\frenchspacing\escapechar=` \input \jobname.ref\vfill\eject}\nonfrenchspacing}
\def\startrefs#1{\immediate\openout\rfile=\jobname.ref\refno=#1}
\def\xref{\expandafter\xr@f}\def\xr@f[#1]{#1}
\def\refs#1{\count255=1[\r@fs #1{\hbox{}}]}
\def\r@fs#1{\ifx\UNd@FiNeD#1\message{reflabel \string#1 is undefined.}%
\nref#1{need to supply reference \string#1.}\fi%
\vphantom{\hphantom{#1}}{\let\hyperref=\relax\xdef\next{#1}}%
\ifx\next\em@rk\def\next{}%
\else\ifx\next#1\ifodd\count255\relax\xref#1\count255=0\fi%
\else#1\count255=1\fi\let\next=\r@fs\fi\next}
%
\newwrite\lfile
{\escapechar-1\xdef\pctsign{\string\%}\xdef\leftbracket{\string\{}
\xdef\rightbracket{\string\}}\xdef\numbersign{\string\#}}
\def\writedefs{\immediate\openout\lfile=\jobname.def \def\writedef##1{%
{\let\hyperref=\relax\let\hyperdef=\relax\let\hypernoname=\relax
 \immediate\write\lfile{\string\def\string##1\rightbracket}}}}%
\def\writestop{\def\writestoppt{\immediate\write\lfile{\string\pageno%
\the\pageno\string\startrefs\leftbracket\the\refno\rightbracket%
\string\def\string\secsym\leftbracket\secsym\rightbracket%
\string\secno\the\secno\string\meqno\the\meqno}\immediate\closeout\lfile}}
\def\writestoppt{}\def\writedef#1{}
\writedefs
\def\biblio\par{\vskip0pt plus.1\vsize\penalty-100\vskip0pt plus-.1
\vsize\bigskip\vskip\parskip
\message{Bibliographie}
{\leftline{\bf \hyperdef\hypernoname{biblio}{bib}{Bibliographical Notes}}}
\nobreak\medskip\noindent\frenchspacing
\writetoca{\string\hyperref{}{biblio}{bib}{Bibliographical Notes}}}%

\def\biblionote{\iffrancmode Notes Bibliographiques\else Bibliographical Notes
\fi}
\def\beginbib\par{\vskip0pt plus.1\vsize\penalty-100\vskip0pt plus-.1
\vsize\bigskip\vskip\parskip
\message{Bibliographie}
{\leftline{\bf \hyperdef\hypernoname{biblio}{\prefix\the\nochapter}%
{\biblionote}}}
\nobreak\medskip\noindent\frenchspacing
\writetoca{\string\hyperref{}{biblio}{\prefix\the\nochapter}%
{\biblionote}}}%

\def\Exercises{\iffrancmode Exercices\else Exercises
\fi}
\def\exerc\par{\vskip0pt plus.1\vsize\penalty-100\vskip0pt plus-.1
\vsize\bigskip\vskip\parskip\global\noexerc=0
\message{Exercises}
\iffrancmode\mark{Exercices}\else\mark{Exercises}\fi
{\leftline{\bf\hyperdef\hypernoname{exercise}{\the\nochapter}{\Exercises}}}
\nobreak\medskip\noindent\frenchspacing
\writetoca{\string\hyperref{}{exercise}{\the\nochapter}{\Exercises}}
}
\def\esubsec{\ifnum\noexerc=0\vskip-12pt\else\vskip0pt plus.05\vsize%
\penalty-100\vskip0pt plus-.05\vsize\bigskip\vskip\parskip\fi%
\global\advance\noexerc by 1
\hyperdef\hypernoname{exercise}{\the\nochapter.\the\noexerc}{}%
\vbox{\noindent\it \iffrancmode Exercice\else Exercise\fi\ \the\nochapter.\the\noexerc}\smallskip\noindent}
\def\exelbl#1{\ifdraftmode{\hfill\escapechar-1{\rlap{\hskip-1mm%
\sevenrm\string#1}}}\fi%
{\xdef#1{\noexpand\hyperref{}{exercise}{\the\nochapter.\the\noexerc}%
{\the\nochapter.\the\noexerc}}}%
\edef\ewrite{\write\equa{{\string#1}\the\nochapter.\the\noexerc}%
\write\equa{}}\ewrite%
\writedef{#1\leftbracket#1}}

\def\eqnn{\global\advance\neqno by 1 \ifinner\relax\else%
\eqno\fi(\prefix\the\nochapter.\the\neqno)}
%
\def\eqnd#1{\DefWarn#1%
\global\advance\neqno by 1
{\xdef#1{($\noexpand\hyperref{}{equation}{\prefix\the\nochapter.\the\neqno}%
{\prefix\the\nochapter.\the\neqno}$)}}
\ifinner\relax\else\eqno\fi(\hyperdef\hypernoname{equation}{\prefix\the%
\nochapter.\the\neqno}{\prefix\the\nochapter.\the\neqno})
\writedef{#1\leftbracket#1}
\ifdraftmode{\escapechar-1{\rlap{\hskip.2mm\sevenrm\string#1}}}\fi
\edef\ewrite{\write\equa{{\string#1},(\prefix\the\nochapter.\the\neqno)
{\noexpand\number\pageno}}\write\equa{}}\ewrite}
%
\def\checkm@de#1#2{\ifmmode{\def\f@rst##1{##1}\hyperdef\hypernoname{equation}%
{#1}{#2}}\else\hyperref{}{equation}{#1}{#2}\fi}
\def\f@rst#1{\c@t#1a\em@ark}\def\c@t#1#2\em@ark{#1}
\def\eqna#1{\DefWarn#1%
\global\advance\neqno by1\ifdraftmode{\hfill%
\escapechar-1{\rlap{\sevenrm\string#1}}}\fi%
\xdef #1##1{(\noexpand\relax\noexpand%
\checkm@de{\prefix\the\nochapter.\the\neqno\noexpand\f@rst{##1}1}%
{\hbox{$\prefix\the\nochapter.\the\neqno##1$}})}
\writedef{#1\numbersign1\leftbracket#1{\numbersign1}}%
}
\def\em@rk{\hbox{}}
\def\xeqn{\expandafter\xe@n}\def\xe@n(#1){#1}
\def\xeqna#1{\expandafter\xe@na#1}\def\xe@na\hbox#1{\xe@nap #1}
\def\xe@nap$(#1)${\hbox{$#1$}}
\def\eqns#1{(\e@ns #1{\hbox{}})}
\def\e@ns#1{\ifx\UNd@FiNeD#1\message{eqnlabel \string#1 is undefined.}%
\xdef#1{(?.?)}\fi{\let\hyperref=\relax\xdef\next{#1}}%
\ifx\next\em@rk\def\next{}%
\else\ifx\next#1\xeqn#1\else\def\n@xt{#1}\ifx\n@xt\next#1\else\xeqna#1\fi
\fi\let\next=\e@ns\fi\next}
\def\figure#1#2{\global\advance\nofigure by 1 \vglue#1%
\hyperdef\hypernoname{figure}{\the\nofigure}{}%
{\elevenpoint
\setbox1=\hbox{#2}
\ifdim\wd1=0pt\centerline{Fig.\ \the\nofigure\hskip0.5mm}%
\else\def\caption{Fig.\ \the\nofigure\quad#2\hskip0mm}
\setbox0=\hbox{\caption}
\ifdim\wd0>\hsize\noindent Fig.\ \the\nofigure\quad#2\else
                 \centerline{\caption}\fi\fi}\par}
\def\lfigure#1#2{\global\advance\nofigure by
1\vglue#1%
\hyperdef\hypernoname{figure}{\the\nofigure}{}%
\leftline{\elevenpoint\hskip10truemm  Fig.\
\the\nofigure\quad #2}}
\def\figlbl#1{\ifdraftmode{\hfill\escapechar-1{\rlap{\hskip-1mm%
\sevenrm\string#1}}}\fi%
{\xdef#1{\noexpand\hyperref{}{figure}{\the\nofigure}%
{\the\nofigure}}}%
\edef\ewrite{\write\equa{{\string#1}\the\nofigure}%
\write\equa{}}\ewrite%
\writedef{#1\leftbracket#1}}
\def\tablbl#1{\global\advance\notable by
1\ifdraftmode{\hfill\escapechar-1{\rlap{\hskip-1mm%
\sevenrm\string#1}}}\fi%
\hyperdef\hypernoname{table}{\the\notable}{}
{\xdef#1{\noexpand\hyperref{}{table}{\the\notable}%
{\the\notable}}}%
\edef\ewrite{\write\equa{{\string#1}\the\notable}%
\write\equa{}}\ewrite%
\writedef{#1\leftbracket#1}}

\catcode`@=12




\tolerance 10000

\nref\Sundborg{
  B.~Sundborg,
  ``Stringy gravity, interacting tensionless strings and massless higher spins,''
 Nucl.\ Phys.\ Proc.\ Suppl.\  {\bf 102}, 113 (2001).
 [hep-th/0103247].
 }

\nref\Sezgin{
  E.~Sezgin and P.~Sundell,
  ``Massless higher spins and holography,''
Nucl.\ Phys.\ B {\bf 644}, 303 (2002), [Erratum-ibid.\ B {\bf 660}, 403 (2003)].
[hep-th/0205131]
~{\sl and}~
JHEP {\bf 0507}, 044 (2005).
[hep-th/0305040].
}
\nref\Klebanov{
  I.~R.~Klebanov and A.~M.~Polyakov,
  ``AdS dual of the critical O(N) vector model,''
Phys.\ Lett.\ B {\bf 550}, 213 (2002).
[hep-th/0210114].}
\nref\Fradkin{
  E.S.~Fradkin ~and~ M.A.~Vasiliev
  ``On the Gravitational Interaction of Massless Higher Spin Fields,''
Phys.\ Lett.\ B {\bf 189}, 89 (1987) {\sl and} M.A.Vasiliev
  ``Higher spin gauge theories: Star product and AdS space,''
In *Shifman, M.A. (ed.): The many faces of the superworld* 533-610.
[hep-th/9910096].
}
\nref\GiombiA{
  S.~Giombi and X.~Yin,
  ``Higher Spin Gauge Theory and Holography: The Three-Point Functions,''
JHEP {\bf 1009}, 115 (2010).
[arXiv:0912.3462 [hep-th]].
~{\sl and }~
  ``Higher Spins in AdS and Twistorial Holography,''
JHEP {\bf 1104}, 086 (2011).
[arXiv:1004.3736 [hep-th]].
~{\sl and }~
  ``On Higher Spin Gauge Theory and the Critical O(N) Model,''
Phys.\ Rev.\ D {\bf 85}, 086005 (2012).
[arXiv:1105.4011 [hep-th]].
}
\nref\GiombiB{
  S.~Giombi, S.~Prakash and X.~Yin,
  ``A Note on CFT Correlators in Three Dimensions,''
JHEP {\bf 1307}, 105 (2013).
[arXiv:1104.4317 [hep-th]].
}
\nref\Javi{
  R.~d.~M.~Koch, A.~Jevicki, K.~Jin and J.~P.~Rodrigues,
  ``$AdS_4/CFT_3$ Construction from Collective Fields,''
Phys.\ Rev.\ D {\bf 83}, 025006 (2011).
[arXiv:1008.0633 [hep-th]] ~{\bf and}~~
``S=1 in O(N)/HS duality,''
Class.\ Quant.\ Grav.\  {\bf 30}, 104005 (2013).
[arXiv:1205.4117 [hep-th]].
}
\nref\Doug{
  M.~R.~Douglas, L.~Mazzucato and S.~S.~Razamat,
  ``Holographic dual of free field theory,''
Phys.\ Rev.\ D {\bf 83}, 071701 (2011).
[arXiv:1011.4926 [hep-th]].
}
\nref\Wadia{
  S.~Giombi, S.~Minwalla, S.~Prakash, S.~P.~Trivedi, S.~R.~Wadia and X.~Yin,
  ``Chern-Simons Theory with Vector Fermion Matter,''
Eur.\ Phys.\ J.\ C {\bf 72}, 2112 (2012).
[arXiv:1110.4386 [hep-th]].
}

\nref\AharonyI{
  O.~Aharony, G.~Gur-Ari and R.~Yacoby,
  ``d=3 Bosonic Vector Models Coupled to Chern-Simons Gauge Theories,''
JHEP {\bf 1203}, 037 (2012).
[arXiv:1110.4382 [hep-th]].
}

\nref\Jain {
  S.~Jain, S.~P.~Trivedi, S.~R.~Wadia and S.~Yokoyama,
  ``Supersymmetric Chern-Simons Theories with Vector Matter,''
JHEP {\bf 1210}, 194 (2012).
[arXiv:1207.4750 [hep-th]].
}

\nref\AharonyII{
  O.~Aharony, G.~Gur-Ari and R.~Yacoby,
  ``Correlation Functions of Large N Chern-Simons-Matter Theories and Bosonization in Three Dimensions,''
JHEP {\bf 1212}, 028 (2012).
[arXiv:1207.4593 [hep-th]].
}

\nref\AharonyIII{
  O.~Aharony, S.~Giombi, G.~Gur-Ari, J.~Maldacena and R.~Yacoby,
  ``The Thermal Free Energy in Large N Chern-Simons-Matter Theories,''
JHEP {\bf 1303}, 121 (2013).
[arXiv:1211.4843 [hep-th]].
}

\nref\Chang{
  C.~-M.~Chang, S.~Minwalla, T.~Sharma and X.~Yin,
  ``ABJ Triality: from Higher Spin Fields to Strings,''
  J.\ Phys.\ A {\bf 46}, 214009 (2013)
  [arXiv:1207.4485 [hep-th]].
}
\nref\MosZinn{
  M.~Moshe and J.~Zinn-Justin,
  ``Quantum field theory in the large N limit: A Review,''
Phys.\ Rept.\  {\bf 385}, 69 (2003).
[hep-th/0306133].
}

\nref\BMB{
  W.~A.~Bardeen, M.~Moshe and M.~Bander,
  ``Spontaneous Breaking of Scale Invariance and the Ultraviolet
  Fixed Point in O($N$) Symmetric $\Phi^{6}$ in Three-Dimensions Theory,''
Phys.\ Rev.\ Lett.\  {\bf 52}, 1188 (1984).
}
\nref\Rabino{
  D.~J.~Amit and E.~Rabinovici,
  ``Breaking of Scale Invariance in $\phi^6$ Theory: Tricriticality and Critical End Points,''
Nucl.\ Phys.\ B {\bf 257}, 371 (1985)..
}
\nref\Bardeen{
  W.~A.~Bardeen, K.~Higashijima and M.~Moshe,
  ``Spontaneous Breaking of Scale Invariance in a Supersymmetric Model,''
Nucl.\ Phys.\ B {\bf 250}, 437 (1985).
}

\nref\Feinberg{
  J.~Feinberg, M.~Moshe, M.~Smolkin and J.~Zinn-Justin,
  ``Spontaneous breaking of scale invariance and supersymmetric models at finite temperature,''
Int.\ J.\ Mod.\ Phys.\ A {\bf 20}, 4475 (2005). }
\vskip -1.5cm

June 10 2014

\preprint{CERN-PH-TH/2013-299 \ \ \ \ }
\vskip -2cm
\preprint{FERMILAB-PUB-13-552-T}

\bigskip

\bigskip
\bigskip
\bigskip

\title{Spontaneous Breaking of Scale
Invariance in a D=3 U(N) Model with Chern-Simons Gauge Fields }

\centerline{ {\bf William A. Bardeen}${}^{1,a}$ ~~and~~ {\bf
Moshe~Moshe}${}^{2 ,b}$}
\bigskip
{\baselineskip14pt \centerline{${}^1$\it  Fermilab}
 \centerline{\it  P.O. Box 500. Batavia, IL 60510-5011, USA }
\vskip .3cm

 \centerline{${}^2$\it Department of Physics, Technion
- Israel Institute of Technology,} \centerline{\it Haifa, 32000
ISRAEL}
\smallskip

\footnote{}{(a)~email: bardeen@fnal.gov}
\footnote{}{(b)~email: moshe@technion.ac.il}


\abstract  We study spontaneous breaking of scale
invariance in the large N limit of three dimensional
$U(N)_\kappa$ Chern-Simons theories coupled to a scalar field in
the fundamental representation. When a $\lambda_6
(\phi^\dagger\cdot\phi)^3$ self interaction term is added to the
action we find a massive phase at a certain critical value for a
combination of the $\lambda_6$ and 't Hooft's   $\lambda=N/\kappa$
couplings. This model attracted recent attention since at finite
$\kappa$ it
 contains a singlet
sector which is conjectured
to be dual to Vasiliev's higher spin gravity on $AdS_4$.
Our paper concentrates on the massive phase of the 3d boundary theory.
We discuss the advantage of introducing masses in the boundary theory through
spontaneous breaking of scale invariance.
\endabstract
\vfill\eject

\section ~~~~~~~~~~~~Introduction

 \sslbl\ssA


There is recently new interest in the phase structure of large $N$  ~$O(N)$  and $U(N)$
symmetric theories with matter fields in the fundamental representation.
This interest
\refs{\Sundborg}\refs{\Sezgin}\refs{\Klebanov}
focused on the conjectured
  $AdS/CFT$ correspondence  between the singlet
sector of the $O(N)$
vector theory in d=3 space-time dimensions and Vasiliev's higher spin gravity theory
on $AdS_4$ \refs{\Fradkin}. The conjecture has been also tested by the computation of
correlation functions of the higher spin gauge theory \refs{\GiombiA}\refs{\GiombiB}.
The quantum completion of the gravitational side of this duality is not known.
If only tree level is considered, one has
 in this case an $AdS/CFT$ correspondence for
which both sides are weakly coupled. The advantage, in this case,
is obvious as a testing ground
of $AdS/CFT$ ideas using explicit derivations \refs{\Javi}\refs{\Doug}.

Recent attention was also given to the large N limit of
$O(N)_\kappa$ and $ U(N)_\kappa$ level $\kappa$ Chern-Simons gauge
theories, at $N,\kappa \to \infty$ with a fixed 't Hooft coupling
$\lambda= N/\kappa$, coupled to scalar and fermion matter fields
in the fundamental representation~ \refs{\Wadia-\AharonyIII}. Pure
Chern-Simons gauge theories possess first and second class
constraints resulting in the absence of propagating degrees of
freedom. Thus only the matter fields in the fundamental
representation provide the true canonical degrees of freedom in
these theories.

Interacting scalars and fermions in $O(N)$ and  $U(N)$ symmetric
field theories at large $N$ are well understood  and their phase
structure is known \refs{\MosZinn}. In the presence of the
Chern-Simons gauge field  the singlet sector is singled out and
the system fits well into the $AdS/CFT$ conjectured duality. Using
large N methods and the convenience of the light-cone gauge for
the Chern-Simons action, explicit calculations can be performed
shedding light on the $AdS/CFT$ correspondence. The original
$O(N)$ model of $g(\vec\phi)^4$ in \refs{\Klebanov} was deformed
in \refs{\Wadia}-\refs{\AharonyIII} to include a marginal
interaction term $(\lambda_6/24\pi )(\vec\phi)^6$. It was
conjectured that the gravity dual of these 3d theories on the
boundary  is a parity broken version of Vasiliev's higher spin
theory \refs{\Fradkin} on $AdS_4$ in the bulk with a parity
breaking parameter $\theta$ which depends on the 't Hooft coupling
$\lambda= N/\kappa$. Supersymmetric extension of this idea was
introduced in \refs{\Jain}.

The conjectured  higher spin symmetry is an approximate symmetry valid at large N.
The massless conformal invariant phase
was analysed either by a generalized Hubbard-Stratanovich
method \refs{\Wadia} or in perturbation theory \refs{\AharonyI}.

The theory of scalars and fermions in the fundamental representation of $U(N)$ coupled
to Chern-Simons gauge fields
are dual to the same bulk gravity theory on $AdS_4$. Thus it has also
been conjectured that there is a duality
between the boson and fermion theories \refs{\AharonyII}. Calculation of
the thermal free energy in large N Chern-Simons field coupled to fermions and scalars
further strengthened this duality \refs{\AharonyIII}. The thermal free energy has been calculated also
in \refs{\Wadia}.

Ref. \refs{\AharonyIII} concentrated on conformal theories in their "regular" and
"critical" phase on the d=3 boundary.
It was noted however that in the case of massive fermion and scalars the
thermal free energy can be also calculated and the duality between the two Chern-Simons matter
 theories of fermion and boson
may still exist. In this case  high spin symmetry with the bulk Vasiliev's
$AdS_4$ is lost since the boundary theory is not conformal.

It was also mentioned in Ref. \refs{\AharonyIII} that the introduction of masses through
spontaneous breaking of scale invariance \refs{\BMB-\Feinberg}~is an alternative
at large N. Though at finite $N$ the breakdown of the scale symmetry
is explicit it is of ${\cal O}(N^{-1})$, namely the same order by which the $AdS/CFT$ correspondence
is approximated in the above mentioned conjecture.

We find it therefore appealing to further analyse the Chern-Simons matter theory
with masses introduced through spontaneous breaking of scale invariance
which assures the introduction of masses into the theory
but leaves the boundary d=3 theory
conformal to ${\cal O}(N^{-1})$.

In this paper we are emphasizing the massive phase and spontaneous
breaking of scale invariance that occurs in this model at a fixed
combination of coupling constants, $\lambda^2 + \lambda_6/8\pi^2 =
4$. After a brief introduction in {Section \label{\ssB}}  we
calculate  in {Subsection \label{\ssBa}} the boson mass gap for a
self
 interacting boson in the fundamental representation of $U(N)$
 coupled at a Chern-Simons gauge field. In {Subsection \label{\ssBb}}
the vertex is calculated in the massive phase with spontaneously
broken scale invariance. {Section \label{\ssC}} is devoted to
calculations of several correlation functions in the massive
phase.In {Subsection \label{\ssCa}} we calculate the ladder sum of
the vertex and the resulting two point correlation of the scalar
vertices - the "bubble graph". In {Subsection \label{\ssCb}} the
exact full planar vertex and "bubble graph" is calculated and the
effective four point coupling $\lambda^{eff}_4$ is defined. In
{Subsection \label{\ssCc}}  we calculate the two point correlation
$<J_0 J_0>$ ~in the large $N$ limit  to all orders in 't Hooft
coupling $\lambda= N/\kappa$ and the effective Lagrangian of the
dilaton is introduced. {Subsection \label{\ssCd}}  is devoted to
explicit breaking of scale invariance and the pseudo-dilaton. In
{Subsection \label{\ssCe}}  the three point correlation
is calculated and the dilaton self interaction is
revealed. Summary and conclusions are found in {Section
\label{\ssD}}.


 \par
 \vskip 1.5 cm

\section ~~~Chern-Simons gauge field coupled to a $U(N)$ scalar
- light cone gauge

 \sslbl\ssB

\noindent We will consider  a complex scalar field $\phi(x)$ in
the fundamental representation of U(N) in three Euclidean
dimensions coupled to a Chern-Simons level $\kappa$ gauge field
$A_\mu(x)$. To the free Chern-Simons action
$$\eqalignno{{\cal S}_{\rm CS}({\bf A}) =-{i\kappa\over 4\pi}
\epsilon_{\mu\nu\rho}\int\d^3 x\, Tr\left[ {\bf A}_ \mu(x)
\partial_ \nu  {\bf A}_ \rho(x) + \frac{2}{3}  {\bf A}_\mu(x) {\bf
A}_ \nu(x){\bf A}_\rho(x)\right],~~~~&\eqnd\CSfree\cr}$$ \noindent
we add
 the $U(N)$
invariant action
$${\cal S}_{\rm Scalar}=\int\d^3 x\left[  ({\bf D}_\mu{\phi(x)})^\dagger\cdot
{\bf D}_\mu{\phi(x)} +NV({\phi(x)}^\dagger \cdot{\phi(x)} /N) \right],  \eqnd\scalarAction$$
where the complex $\phi(x)$ field is in the fundamental representation and ${\bf D}_\mu$
is the covariant derivative
${\bf D}_\mu=\partial_\mu +{\bf A}_\mu\,.$
 Here, $x^\mu=\{x^1,x^2,x^3\}$ ;
$x^\pm ={1 \over \sqrt 2} (x^1\pm i x^2)  $ and
$ {\bf D}_\pm = \partial_\pm +{\bf A}_\pm\, $.
The contributions involving the interaction of the scalar and the
gauge fields are generated by the
covariant  derivatives
$$\eqalignno{  {\bf D}_\mu{\phi}^\dagger\cdot{\bf D}_\mu{\phi}&  =
\partial_3{\phi}^\dagger\cdot\partial_3{\phi}+\partial_+{\phi}^\dagger\cdot
\partial_-{\phi}+\partial_-{\phi}^\dagger\cdot   \partial_+{\phi}\cr& -
{\phi}^\dagger{\bf A}_-\partial_+ {\phi} - {\phi}^\dagger{\bf
A}_+\partial_- {\phi}- {\phi}^\dagger{\bf A}_3\partial_3 {\phi}
\cr& +
\partial_+{\phi}^\dagger{\bf A}_-  {\phi} +
\partial_-{\phi}^\dagger{\bf A}_+ {\phi}  +
\partial_3{\phi}^\dagger{\bf A}_3  {\phi} \cr& -
{\phi}^\dagger\left({\bf A}_3{\bf A}_3+{\bf A}_+{\bf A}_-+{\bf
A}_-{\bf A}_+\right){\phi}  \, \ \ \ \ \ \ \ \ \eqnd\CovDeriv &\cr } $$

\noindent ${\bf A}= A^aT^a$ ~,~  $T^a$  are
antihermitian,
normalized
by $Tr\{T^aT^b\}=-\half\delta^{ab}$. In the light-cone gauge
${\bf A}_-={1\over\sqrt{2}}\left({\bf A}_1 + i {\bf A}_2\right) = 0\, $
and thus
$$\eqalignno{  {\cal S_{\rm CS} + S_{\rm Scalar}}&
=\int\d^3 x \{ {\kappa\over 4\pi}{ A}^a_+\del_-{ A}^a_3
-{\phi}^\dagger (\del^2_3 + 2 \del_+ \del_- )\phi \cr&  -
{\phi}^\dagger{A^a}_+T^a\partial_- {\phi} +
\partial_-{\phi}^\dagger{A^a}_+T^a {\phi}\cr& -
{\phi}^\dagger{A}_3 T^a\partial_3 {\phi}   +
\partial_3{\phi}^\dagger{A^a}_3 T^a {\phi} \cr& -
{\phi}^\dagger\left({ A^a}_3{A^a}_3 T^a T^b \right){\phi}
+NV({\phi}^\dagger \cdot{\phi} /N) \} \, \ \ \ \ \ \ \ \eqnd\CSscalar &\cr}
$$
Since in the light-cone gauge the action is linear in $A_+^a$we have simply:
$$-{\kappa\over 4\pi}\partial_- A_3^a  =J^a_- = \phi^\dagger{T}^a\partial_-\phi
- \partial_-\phi^\dagger{T}^a\phi
 \eqnd\EqofMo$$
With proper boundary conditions $A^a_3(x^-,x^+,x^3)$ is given by

$$ A^a_3(x^-,x^+,x^3)={2\pi\over\kappa}\int^\infty_{-\infty}dx'^-\sgn(x'^- - x^-)
J^a_-(x'^-,x^+,x^3)$$
or
$$ A^a_3(p)= ({2\pi\over \kappa}) {2 i p^+\over {p^+}^2+\epsilon^2}J^a_{-}
    \ \ \ \ { _{\epsilon \to 0} }\to
({4 \pi i\over \kappa}){1\over p^+}J^a_-\eqnd\AthreeSol $$

\noindent When inserted in the action, one finds the gauge
field propagator as the principal part of:
$$G_{+3}(p)= -G_{3+}(p)={4\pi i\over \kappa}{1\over p^+}=4\pi i{ \lambda\over N}{1\over p^+}$$
\noindent where the 't Hooft coupling is $\lambda={N\over \kappa}$

The following U(N) invariant self interacting scalar
potential $V({\phi}^\dagger \cdot{\phi} /N)$ will be considered now:
$$NV({\phi}^\dagger \cdot{\phi} /N) =
{\mu^2{\phi}^\dagger \cdot{\phi}}
+
{1\over 2}{\lambda_4\over N}({\phi}^\dagger \cdot{\phi})^2+{1\over 6} {\lambda_6 \over N^2}
({\phi}^\dagger \cdot{\phi})^3 \eqnd\Vphi$$
Following Ref~\refs{\BMB}, we will concentrate on the
scale invariant potential with renormalized
$\mu_R=\lambda_{4R}=0$ and
analyze the system in its massive phase of spontaneously broken conformal invariance.

\vskip 1 cm
\subsection  {\bf ~~~1PI in the Massive Phase with Spontaneously Broken Scale Invariance }

\sslbl\ssBa

\midinsert \epsfxsize=100.mm \epsfysize=55.mm
\centerline{\epsfbox{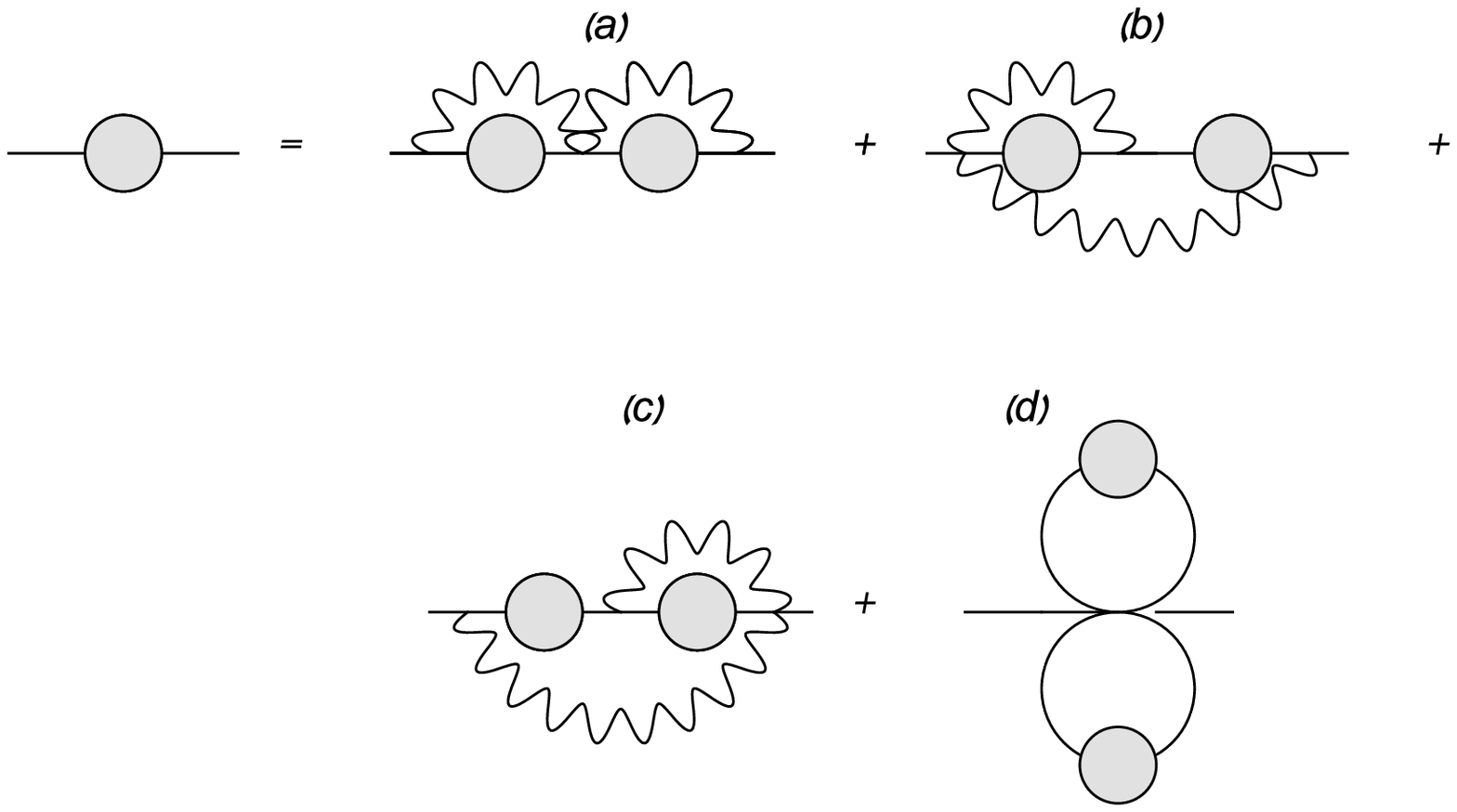}} \figure{3.mm}{1PI self energy
graphs}
\figlbl\selfEnergy
\endinsert

\noindent The one particle irreducible part of the scalar
self-energy is given by the sum of diagrams in Fig.\selfEnergy
~\AharonyIII. The contributions of the diagrams [(a)+(b)+(c)] in
Fig.\selfEnergy ~are:

$$\eqalignno{  {\Sigma^{(a,b,c)}(p,\lambda)_{ij}}&
=\delta_{ij}\int{\d^3q\over (2\pi)^3}\int {\d^3l\over (2\pi)^3}
~\{ \cr&{4\pi^2\lambda^2}
    {(l+p)^+(q+p)^+\over (l-p)^+(q-p)^+}({1\over
(q^2+\Sigma(q))  (l^2+\Sigma(l))}) \cr&
{-8\pi^2\lambda^2}{(l+p)^+(q+l)^+\over (l-p)^+(q-l)^+}({1\over
(q^2+\Sigma(q))  (l^2+\Sigma(l))}) ~~\}  \, &\eqnd\Sigmaabc\cr}
$$
\noindent which sum up to
$$\eqalignno{  {\Sigma^{(a,b,c)}(p,\lambda)_{ij}}&
={4\pi^2\lambda^2}\delta_{ij}\int{\d^3q\over (2\pi)^3}\int
{\d^3l\over (2\pi)^3}{1\over (q^2+\Sigma(q)) (l^2+\Sigma(l))}
&\eqnd\Sigmaabcsum\cr}
$$
The sum of diagrams (a)+(b)+(c) in Eq. \Sigmaabcsum ~is of the same form
as the contribution of diagram (d) of
the scalar self interaction
$$\eqalignno{{\Sigma^{(d)}(p,\lambda)_{ij}}  &
={\half\lambda_6}\delta_{ij}\int{\d^3q\over (2\pi)^3}\int
{\d^3l\over (2\pi)^3}{1\over (q^2+\Sigma(q)) (l^2+\Sigma(l))}
&\eqnd\Sigmad\cr}
$$
It was shown that in
the light-cone gauge $\Sigma (p)$ is a constant \Wadia,\AharonyI ~and thus  the sum of
Eqs. \Sigmaabcsum ~and ~\Sigmad ~is given by:

$$\eqalignno{  {\Sigma(p,\lambda,\lambda_6)}
&={4\pi^2(\lambda^2 + {\lambda_6\over 8\pi^2})\{\int{\d^3q\over
(2\pi)^3}{1\over (q^2+\Sigma(q)) }\}^2 } \cr &= 4\pi^2 (
\lambda^2 + {\lambda_6\over 8\pi^2} ) \{ {1\over 2\pi^2}(\Lambda -
\half\pi {\sqrt |\Sigma|} ) \}^2 &\eqnd \Sigmaabcd }
$$
\noindent A sharp UV cutoff is illustrated in Eq. \Sigmaabcd ~as an example and any
other UV regulator can be employed. The fully renormalized
gap equation can be completed by
 adding to the gap equation the contributions of the
${\mu^2{\phi}^\dagger \cdot{\phi}}$
~and~
${1\over 2}{\lambda_4\over N}({\phi}^\dagger \cdot{\phi})^2$
 terms from Eq. \Vphi.
Namely, on the right hand side of Eq.\Sigmaabcd will be added:

$$\lambda_4\int{\d^3q\over
(2\pi)^3}{1\over (q^2+\Sigma(q)) } +\mu^2
 $$

The renormalized gap equation will be now:
$$\eqalignno{  {\Sigma(p,\lambda,\mu,\lambda_4\lambda_6)}
 &= {1\over 4}( \lambda^2 + {\lambda_6\over 8\pi^2} )|\Sigma| -
 \lambda_{4R}{{\sqrt |\Sigma|}\over 4\pi}
 +\mu^2_R &\eqnd \SigmaRen }
$$

\noindent In the conformal phase \BMB
$$\eqalignno{ \mu_R=\lambda_{4R}=0~~~~~~~~~~~~~~~~~~ \ \ \ \
\ \ \ \ \ \ \ \ \ \ \ \ \ \ \ \ \ \   \eqnd \confPhase }$$
The renormalized 1PI
self-energy is given by the solution of the gap equation :
$$ \eqalignno{\Sigma
= {1\over 4}( \lambda^2 + {\lambda_6\over 8\pi^2} )|\Sigma|
&&\eqnd \gapEq }$$

Namely, there are two possible solutions:

\noindent (a) $\Sigma =M^2= 0$ gives the conformal invariant massless phase discussed
in Refs.\refs{\Wadia}-\refs{\AharonyII}

or

\noindent (b) $\Sigma=M^2 \neq  0$ ~~if~~ $  \lambda^2 + {\lambda_6\over 8\pi^2}
= 4 $  which results in a massive phase for this critical combination
of the 't Hooft $\lambda$ and self-interacting $\lambda_6 $  couplings. This is a
spontaneously broken scale invariance phase similar to the one encountered in Ref. \refs\BMB
~in
the case of the O(N) invariant self interacting $(\phi \cdot\phi)^3$ theory
in three dimensions.
This relation can be also written as
$$\eqalignno{\lambda^2 + {\lambda_6\over 8\pi^2}
= \lambda^2 + 4{\lambda_6\over \lambda_6^{crt}}
=4 &&\eqnd \critical }$$
where $\lambda_6^{crt}= 32\pi^2$ is the critical value of the pure U(N) invariant
case and is analogous to  $\lambda_6^{crt}= 16\pi^2$ in the case of the pure O(N)
invariant
$(\phi\cdot\phi)^3$ case \BMB.

\noindent At the critical combination of the couplings in
Eq.\critical ~the massive phase is continually connected to the
massless phase. This degeneracy, as seen also in the variational
calculation in Ref.\refs\BMB~, is due to a flat direction in the
ground state energy. The degeneracy can be lifted by adding soft
breaking terms and a unique ground state for either phase could be
then created. A light pseudo-Goldstone particle should appear in
the spectrum of the massive phase.  One expects that the symmetric
and broken phases will be degenerate in the conformal limit of the
theory. Below the critical coupling the massless conformal phase
is unique without the addition of soft breaking terms.

\vskip .3 cm
\subsection  {\bf ~~~~~~~~~~~~~~~~The vertex in the massive phase at order $\lambda$}

\sslbl\ssBb

\vskip .5cm

\midinsert
\epsfxsize=20.mm
\epsfysize=15.mm
\centerline{\epsfbox{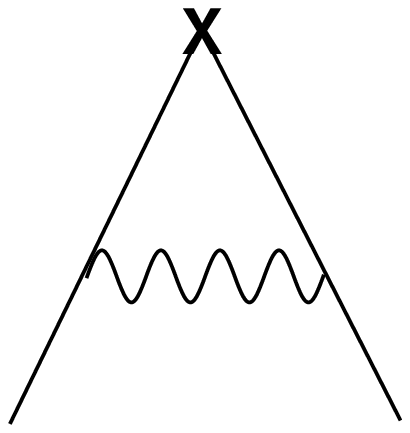}}
\figure{3.mm}{Vertex to order $\lambda$}
\figlbl\vertexlambda
\endinsert

The vertex in Fig. \vertexlambda
~~at $k_+ = 0 $ to
first order in $\lambda$ is given by:
\vskip .1cm
$$\eqalignno{  {V(p^2,k_3)}
&={1+{4\pi i \over \kappa}} \int {d^3 l \over (2\pi)^3}{1\over
(l-p)^+}\cr & \{iT^a((p+l)^+(p+l+2k)^3 -(p+l)^3(p+l+2k)^+ )iT^a\}
\cr& {1\over l^2+\Sigma} {1\over (l+k)^2+\Sigma} \cr & ={1+{i 4\pi
\lambda}} k_3 \int {d^3 l \over (2\pi)^3}{(l+p)^+\over (l-p)^+}
{1\over l^2+\Sigma} {1\over
(l+k)^2+\Sigma}\cr &&\eqnd\orderlambda }
$$

After the angular integration in the 2d $l$ plane

$$\eqalignno{  \int^{2\pi}_0 {d\phi_l \over 2\pi}{(l+p)^+\over(l-p)^+}=
\Theta(l^2-p^2)-\Theta(p^2-l^2)=\epsilon(l^2-p^2) \ \ \ \ \ \ \ \
 \eqnd\angleIntegration}
$$
\noindent where $l^2$ and $p^2$ denote the variables in the two dimensional space,
and after integrating also on $l_3$~, ~$V(p^2,k_3)$ is given by:

$$\eqalignno{  {V(p^2,k_3)}
&= {1+{i \lambda k_3\over 4} \int^\infty_0 d l^2}
\epsilon(l^2-p^2)\int_0^1 dx (l^2+x(1-x)k_3^2+M^2)^{-3/2} \cr &=1+
{i \lambda k_3\over 2}\int_0^1 dx \{2(p^2+x(1-x)k_3^2+M^2)^{-1/2}
-(x(1-x)k_3^2+M^2)^{-1/2}\} \cr &=1+i\lambda~\{
~2~\arctan({k_3\over 2\sqrt{p^2+M^2}}) -\arctan({k_3\over 2M})~\}
~~~~~~~~~~~~~~~~~~~~~~~\eqnd\orderlambdaB }
$$

\noindent One notes that at momentum transfer
$k_+ =0$  ~~$V(p^2,k_3)$ depends only on the two
dimensional vector $\vec p$ and on the
momentum
transfer $k_3$~.

\vskip .3cm
\subsection  {\bf ~~~~The vertex in the massive phase, "seagull" graph contributions}

\sslbl\ssBc

As seen in Eqs. \Sigmaabcsum ~and ~\Sigmad ~,
the two
contributions,~$\lambda^2$ ~and~ $\lambda_6$~,
result
in
similar
contributions
to the
self-energy $\Sigma$~.
For the same reason when the vertex is calculated by adding an
insertion at the appropriate lines in Fig. 1, one finds again that
the contribution to the vertex of order $\lambda^2$
are of the same form as the order
$\lambda_6$ contribution.

\midinsert \epsfxsize=100.mm \epsfysize=55.mm
\centerline{\epsfbox{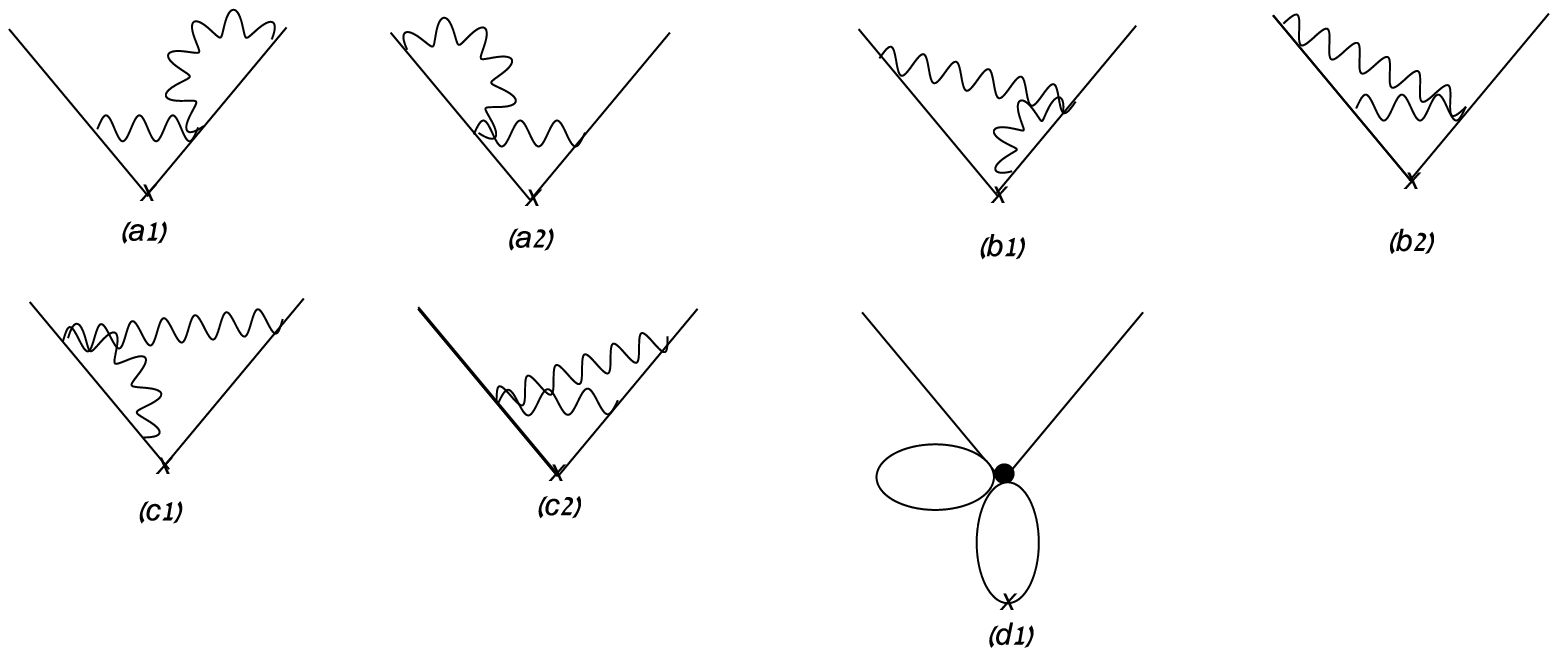}} \figure{3.mm}{"Seagull"
contributions to the vertex diagrams}
\figlbl\Seagull
\endinsert

\noindent One finds the contribution to the vertex
of diagrams a1-2, b1-2 and c1-2 in Fig.\Seagull
$$\eqalignno{  { V^{(a1-2,b1-2,c1-2)}(p,k)}
&={ \lambda^2}4\pi^2 \int {d^3 l \over (2\pi)^3}{d^3 q
\over (2\pi)^3}  \ \ \ \
\cr
& \{ ~ ~-({l+p\over l-p}){^+}({q+p+k\over q-p})^+ ({1\over
l^2+\Sigma}) ({1\over (l+k)^2+\Sigma})({1\over (q+k)^2+\Sigma})
\cr & -({l+p\over l-p})^+({q+p+k\over q-p})^+ ({1\over
l^2+\Sigma}) ({1\over q^2+\Sigma})({1\over (q+k)^2+\Sigma}) \cr &
\cr & +({l+p\over l-p})^+({q+l+2k\over q-l})^+ ({1\over
l^2+\Sigma}) ({1\over (l+k)^2+\Sigma})({1\over (q+k)^2+\Sigma})
\cr & +({l+p\over l-p})^+({l+q\over q-l})^+ ({1\over l^2+\Sigma})
({1\over q^2+\Sigma})({1\over (q+k)^2+\Sigma})\cr &  +({l+q+k\over
l-q})^+({q+p+2k\over q-p})^+ ({1\over l^2+\Sigma}) ({1\over
(l+k)^2+\Sigma})({1\over (q+k)^2+\Sigma}) \cr & +({l+q\over
l-q})^+({q+p+k\over q-p})^+ ({1\over l^2+\Sigma}) ({1\over
q^2+\Sigma})({1\over (q+k)^2+\Sigma}) ~\} \ \ \ \ \ \ \ \ \eqnd\vertexabc}
$$
\noindent The self interaction of the scalar fields contributes to the vertex the term
$$\eqalignno{  {V^{(d1-2)}(p,k)}
&={-\half \lambda_6} \int {d^3 l \over (2\pi)^3}{d^3 q \over
(2\pi)^3} \ \ \ \
\{ ~  ({1\over
l^2+\Sigma}) ({1\over (l+k)^2+\Sigma})({1\over q^2+\Sigma}) \cr &
\ \ \ \ \ +({1\over l^2+\Sigma}) ({1\over q^2+\Sigma})({1\over
(q+k)^2+\Sigma})  ~\}&\eqnd\vertexd }
$$

\noindent When all vertex contributions are added at $k^+ = 0 $,
diagrams ~~a1-2, ~~b1-2, ~~c1-2,~~d1-2 result in:

$$\eqalignno{  {V^{(a-d)}(p,k_3)}= V
&=- 8\pi^2({\lambda^2 +{\lambda_6\over 8\pi^2}}) \int {d^3 l \over
(2\pi)^3}({1\over l^2+\Sigma})\int {d^3 q \over (2\pi)^3} \ \ \ \
\cr & ~~~~~~~~~~~~~({1\over (l+k)^2+\Sigma})({1\over q^2+\Sigma})
 &\eqnd\vertexabcdsum}
$$

\noindent Namely, the combined contribution of order $\lambda^2$ and $\lambda_6$ results
in a local vertex. The mutual cancelations which are presented in Eqs.\Sigmaabcd
~and~ \vertexabcdsum
~is a general property of
this model and occurs also in other amplitudes.
We notice the coefficient $-8\pi^2({\lambda^2 +{\lambda_6\over 8\pi^2}})$
equals $-32\pi^2$ at the critical coupling for the massive phase.

\vskip .3cm
\section ~~~~~~~~~~~~~~~~Correlations

\sslbl\ssC

\subsection {\bf ~~~~~~~~Vertex and "bubble" graph  at leading $N$}

\sslbl\ssCa
\vskip .2cm

\noindent At large N and $k^+=0$, the vertex can be calculated
 to all orders in 't Hooft's
coupling constant $\lambda = {N \over \kappa}$. The order
$\lambda$ vertex corrections can be iterated, at $k^+ = 0$ ~by
summing ladders to all orders as shown in Fig. 4.

\midinsert
\epsfxsize=100.mm
\epsfysize=45.mm
\centerline{\epsfbox{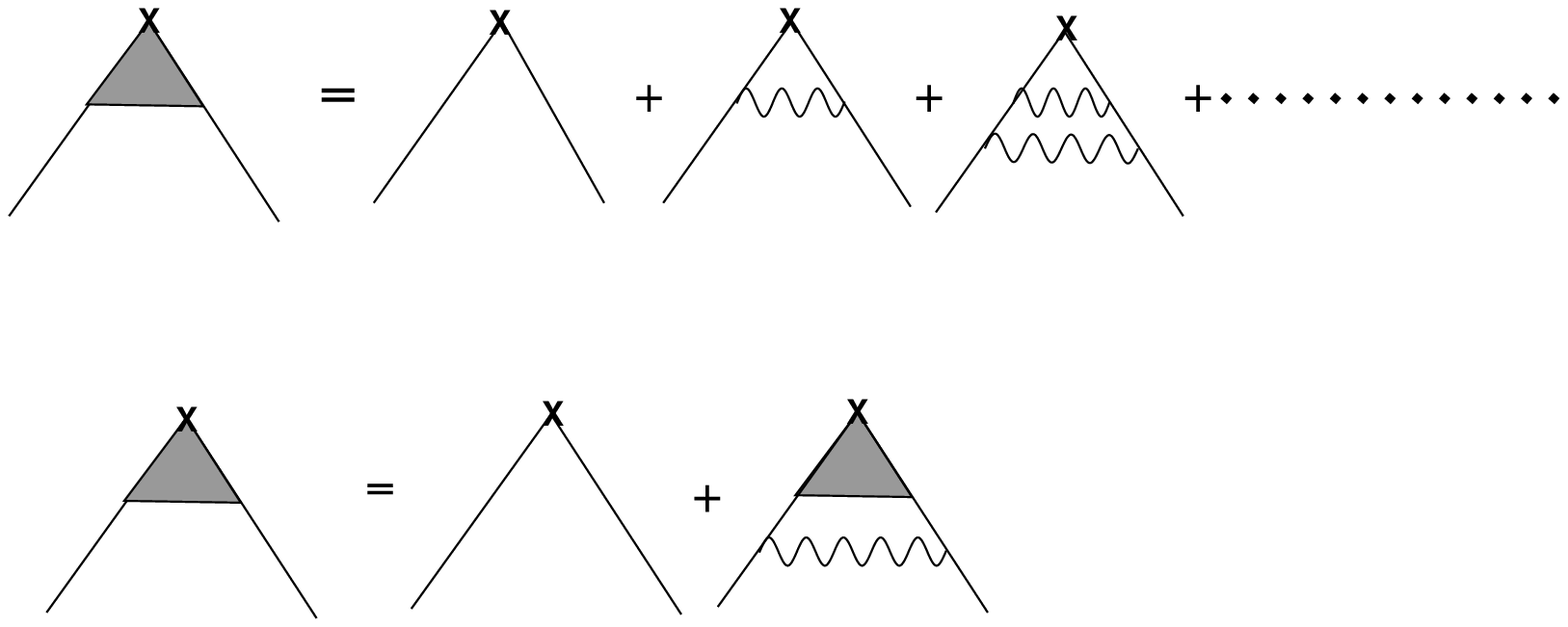}}
\figure{3.mm}{Vertex  in the ladder approximation}
\figlbl\FigVertexLadder
\endinsert

\noindent The resumed vertex satisfies the following integral
equation which can be explicitly evaluated
$$\eqalignno{  {V(p^2,k_3)}
&={1+{4\pi i \over \kappa}} \int {d^3 l \over
(2\pi)^3}V(l^2,k_3){1\over (l-p)^+}\cr & \{iT^a((p+l)^+(p+l+2k)^3
-(p+l)^3(p+l+2k)^+ )iT^a\} \cr& {1\over l^2+\Sigma} {1\over
(l+k)^2+\Sigma} \cr & ={1+{i 4\pi \lambda}} k_3 \int {d^3 l \over
(2\pi)^3}V(l^2,k_3){(l+p)^+\over (l-p)^+}
{1\over l^2+\Sigma} {1\over
(l+k)^2+\Sigma}\cr &&\eqnd\vertexLadder }
$$

After the angle integration in the 2d $l$ plane using Eq. \angleIntegration
~and after integrating also on $l_3$ we have:

$$\eqalignno{  {V(p^2,k_3)}
&= {1+{i \lambda k_3\over 4} \int^\infty_0 d l^2}
\epsilon(l^2-p^2)V(l^2,k_3)\int_0^1 dx (l^2+x(1-x)k_3^2+M^2)^{-3/2}
\cr &&\eqnd\vertexLadderA }
$$
\noindent where $l^2$ and $p^2$ denote the
variables in the two dimensional transverse space.
The general solution for $V(p^2,k_3)$ takes the form

$$\eqalignno{  {V(p^2,k_3)}
&= C \exp \{i \lambda k_3 \int dx (p^2+x(1-x)k_3^2+M^2)^{-1/2}\}
&\eqnd\Vofpk }
$$
\noindent and the constant C is found using Eq.\vertexLadderA
~and therefore satisfies.

$$ C+ V(p^2=0,k_3) = 2 $$

Finally, we have for the vertex function

$$\eqalignno{  {V(p^2,k_3)}
&= 2 \exp \{i \lambda k_3 \int_0^1 dx (p^2+x(1-x)k_3^2+M^2)^{-1/2}\}
\cr & \{  1
+ \exp [i\lambda k_3\int_0^1 (x(1-x)k_3^2+M^2)^{-\half}]\}^{-1}
\cr &&\eqnd\vertexLadderfinal }
$$

\noindent The ladder contribution to the two point correlation of
the scalar vertices can now be computed at $k^+=0$. We denote this
by the  "bubble graph"  $B_{CS}(k_3)$ ~depicted in Fig. 5

\midinsert \epsfxsize=100.mm \epsfysize=10.mm
\centerline{\epsfbox{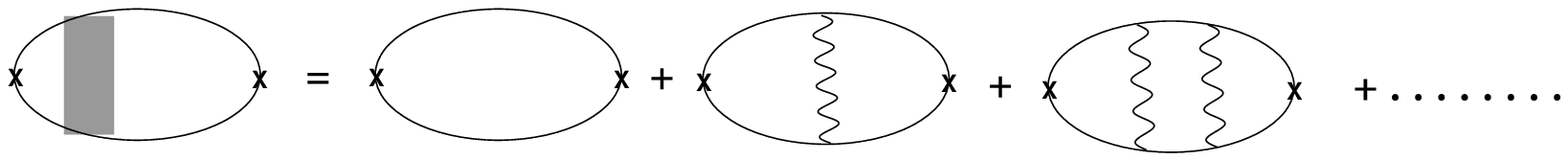}} \figure{3.mm}{$B_{CS}(k_3)$ The
"bubble graph" $B_{CS}(k_3)$} \figlbl\bubbleGraph
\endinsert

$$\eqalignno{ B_{CS}(k_3)
&= \int {d^3 l \over (2\pi)^3} V(l^2,k_3) ({1\over
(l^2+l_3^2+\Sigma})({1\over l^2+(l_3+k_3)^2+\Sigma}) \cr & ={1
\over 16 \pi}\int dl^2 V(l^2,k_3)\int dx
(l^2+x(1-x)k_3^2+\Sigma)^{-3/2} &\eqnd\BCS }
$$

Using the result for the vertex in Eq. \vertexLadderfinal ,  the
integral equation in Eq. \BCS ~gives the explicit expression for $B_{CS}(k_3)$ :

$$\eqalignno{ B_{CS}(k_3)
&={1\over 4\pi \lambda}~{1\over k_3}\tan\{\half \lambda k_3\int dx
(x(1-x)k_3^2+M^2)^{-1/2} \} \cr &= {1\over 4\pi \lambda}~{1\over
k_3}\tan\{ \lambda~\arctan({k_3\over 2M})\} &\eqnd\BCSfinal }
$$

\noindent where $\Sigma=M^2$.

\vskip .2cm

\subsection {\bf ~~~~~~~~~~~~~~~The exact full planar vertex and "bubble graph" }

\sslbl\ssCb
\vskip .2cm

To compute the full planar scalar vertex, we need to be able to
 sum the contributions of the "seagull" graph exchanges
 combined with the ladder diagrams.
 At lowest order the "seagull" exchanges reduce to
 a bubble correction analogous to the insertion of
 the local~ $\lambda_6$ ~vertex contribution
 as in Fig. \Seagull (d) with
 the effect of simply renormalizing ~$\lambda_6$ ~by
 a~ $\lambda^2$ ~term. Ladder
 corrections to the "seagull" vertex are therefore
 the same as for a bare vertex.

 The "seagull" corrections to the ladder vertex are
 more subtle as we must be able
 to perform the angular integrations that
 reduce the "seagull" exchange contributions to
 a local vertex as in  Sec. \ssBc .
 ~Namely, one should be
able to show that the local expression is obtained also when the
lower vertex in Fig.\Seagull ~is not a constant but a function of the loop
variables $l_3$ and $l^2$ , the two dimensional perpendicular variable.
Indeed, by inspecting
Eqs. \vertexabc ~one recognizes two similar observations
that show that this is clearly the case when $k+=0$.

\noindent (a)~Integrating the angle in all terms in Eq. \vertexabc
~using Eq. \angleIntegration
~and shifting the $l_3$ integration by $k_3$ when needed
one notes that all the terms in front of the scalar propagators
in Eq. \vertexabc ~sum up to a constant.

\noindent (b) ~Similarly, by inserting $k_+=0$ in Eq. \vertexabc ~and
shifting the $l_3$
integration by $k_3$ when needed, one finds again that the six expressions in front of
the scalar propagators add up to a constant,
\vskip .5cm
\midinsert
\epsfxsize=120.mm
\epsfysize=10.mm
\centerline{\epsfbox{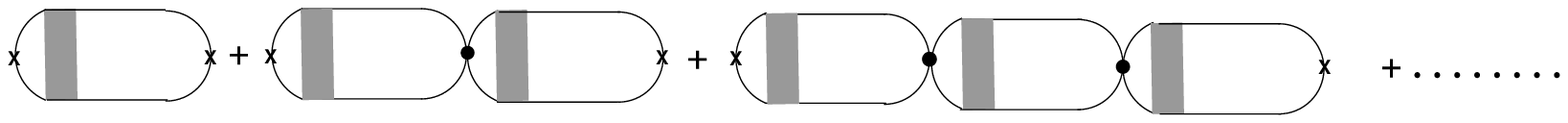}}
\figure{6.mm}{Full planar "bubble graph" }
\figlbl\planarbubble
\endinsert
\vskip .5cm

\midinsert
\epsfxsize=120.mm
\epsfysize=10.mm
\centerline{\epsfbox{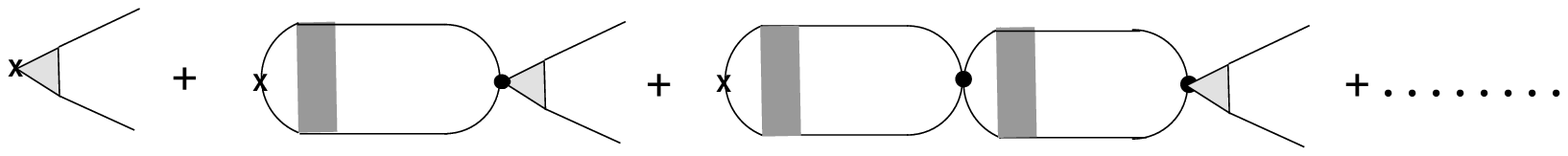}}
\figure{3.mm}{Full planar vertex }
\figlbl\planarVertex
\endinsert

\vskip .2cm

In order to obtain the full scalar vertex we must sum to all orders
iterations of the ladder and "seagull" exchanges. This
results in
a resummation of the "bubble" contributions depicted in Fig. \planarbubble
~and Fig.\planarVertex. Hence, at leading order in the large N expansion, the scalar
vertex at $k^+=0$ is given by

$$\eqalignno{ W(p^2,k_3)&=V(p^2,k_3) - B_{CS}(k_3)\lambda^{eff}_4
V(p^2,k_3) \cr & \ \ \ \ \ + B_{CS}(k_3)\lambda^{eff}_4 V(p^2,k_3)
B_{CS}(k_3)\lambda^{eff}_4 V(p^2,k_3) + \ .\ .\ .  \cr &=
V(p^2,k_3) - B_{CS}(k_3)\lambda^{eff}_4 W(p^2,k_3)\cr &=V(p^2,k_3)
(1+ \lambda^{eff}_4 B_{CS}(k_3))^{-1} &\eqnd\WfullLadder }
$$
\noindent where

$$\eqalignno{ \lambda^{eff}_4=\lambda_4+(\lambda^2
+{\lambda_6\over 8\pi^2}){8\pi^2\over N} <\phi\dag\cdot \phi>
&&\eqnd\eff }$$

\noindent $\lambda_4$ is the self coupling $ {1\over 2}
{\lambda_4\over N}(\phi\dag\cdot\phi)^2$ in Eq. \Vphi ~and

$$ \eqalignno{  <\phi\dag\cdot \phi> = N \int {d^3q \over (2\pi)^3 }
{1 \over q^2 + M^2} = N\{ {\Lambda \over 2\pi^2}- {M\over 4\pi}
\}\ \ \ \ \ \ \ \ \ &&~~~~~~~~~\eqnd\cutoff }
$$

In the absence of explicit breaking of conformal symmetry

 $$\eqalignno{ \lambda_{4R}=\lambda_4 +
 4 \Lambda(\lambda^2+{\lambda_6\over 8\pi^2})=0
 &&~~~~~~~~~\eqnd\hlambdaR }$$

In the massless symmetric phase
$$\lambda^{eff}_4=0$$

 \noindent However, the effective coupling $\lambda^{eff}_4$ in the massive
 spontaneously broken phase is:
$$\lambda^{eff}_4=\lambda_{4R}-2\pi M(\lambda^2+{\lambda_6\over 8\pi^2})=-8\pi M
\eqnd\lambdaEFF$$
\vskip 1cm

\noindent We summarize our results for the
scalar vertex in the massive phase which
are exact to all orders in the ‘t Hooft coupling, $\lambda$~,
and the $\phi^6$~ coupling, $\lambda_6$~, and
to the leading order of the large $N$ expansion.
In this limit, the full scalar vertex is
$$\eqalignno{ W(p^2,k_3)=V(p^2,k_3) (1- 8\pi M B_{CS}(k_3))^{-1}
&&\eqnd\WfullLadderSummary }
$$

\noindent where $V(p^2,k_3)$ is the ladder vertex of
Eq.\vertexLadderfinal
and $B_{CS}(k_3)$ is the bubble function given in Eq.\BCSfinal


\noindent  The above formulas reduce to those of ref. \refs{\AharonyII}
in the massless limit, $M \to 0$,
corresponding to the phase with explicit conformal symmetry.

\subsection {\bf  ~~~~~~~~~~~$<J_0 J_0>$ correlator and the dilaton}

\sslbl\ssCc
 \vskip .2cm

The full correlator
of
two
scalar
currents
at
large
$N$
is
now
simply
computed
from
the
bubble
integral
with
one
bare
vertex,
one
full
vertex
and
the
full
scalar
propagators.
The resulting $<J_0(k) ~J_0(-k)>$  correlator
~is
analogous
to
the
result
found
in
pure $(\vec \phi^2)^3$
theory \refs \BMB

$$<J_0(k) ~J_0(-k)>~=~ N~B_{CS}(k_3)
(1+\lambda_4^{eff}B_{CS}(k_3))^{-1}\eqnd\BCSdef$$

\noindent The
scalar
currents
are
gauge
invariant
so
we
expect
the
scalar
correlator
to
be
a
function
of $\vec k^2$
although
the
explicit
calculation
was
done
with $k^+=0$ .

The bubble summation, explicit in Eq.\BCSdef, allows for the
possibility of poles in the scalar correlator which would signify
the existence of a composite dilaton. In the absence of explicit
breaking of the conformal symmetry, the massive phase must contain
a massless scalar dilaton.

The bubble function, $B_{CS}(k^2)$ in Eq.\BCSdef ~can be written
as


$$\eqalignno{ B_{CS}(k)&=
 {1\over
8\pi M}~+
 {(\lambda^2-1)\over 24\pi M }({k\over 2 M})^2
\{  1 + ({k\over {2 M}})^2 ({2\lambda^2-3\over 5}) \cr & +({k\over
{2 M}})^4 ({17\lambda^4-53\lambda^2+45\over 105}) + ({k\over {2
M}})^6 ({62\lambda^6-295\lambda^4+503\lambda^2-315\over 945}) \cr
& + . \ . \ .\ . \ . \cr & = {1\over 8\pi M}~+
 {(\lambda^2-1)\over 24\pi M }({k\over 2 M})^2
\sum_0^\infty ({k\over {2 M}})^{2n} P_{2n}(\lambda)
&\eqnd\BCSexpan}
$$
 One notes that at $\lambda^2=1$ we have $B_{CS}(k)=
 {1
/8\pi M}$ for all values of $k$. Thus, in the conformal limit
where $ \lambda_4^{eff} = - 8\pi M $ it is clearly seen that the
pole at $\lambda^2=1$ in Eq. \BCSdef ~determines  the physical
boundary of $\lambda$.

Using the expansion in Eq. \BCSexpan ~, the low momentum behavior
of the scalar correlator becomes

$$\eqalignno{ <J_0(k) ~J_0(-k)>~&={N\over 8\pi M }\{1-{k^2\over 12M^2}
(1-\lambda^2)+ . . .\} \cr &~~~~\{1+({\lambda_4^{eff}\over 8\pi M})
(1-{k^2\over 12M^2}(1-\lambda^2)+ . . .) \}^{-1}
&\eqnd\JJexpanded }
$$
At $ \lambda_4^{eff} = - 8\pi M $ the constant terms in the
denominator cancel leaving the massless dilaton pole
$$\eqalignno{ <J_0(k) ~J_0(-k)>~&={3N\over 2\pi  }({M\over 1-\lambda^2})
{1\over k^2} ={f_D^2\over k^2} &\eqnd\dilatonPole }
$$
\noindent where
$$\eqalignno{  f_D=\sqrt{{3N M \over 2\pi(1-\lambda^2)}}
&&\eqnd\fD }$$ \noindent Note the pole in $f_D^2$ at $\lambda^2 =
1$ which is, as mentioned, the boundary of physical couplings for
$\lambda$.

\noindent From the expression for the full scalar vertex function
the residue of the pole yields the coupling of the dilaton to
scalar particles in the effective Lagrangian of the dilaton
interactions. In terms of the dilaton field $D(x)$ ~(where
$J_0(x)=f_D D(x)$) the effective Lagrangian is given by:
$$\eqalignno{ {\cal L} &=\half{ \del}_\mu D\cdot{ \del}_\mu D-g_D
(\phi^\dagger \cdot \phi ) D
&\eqnd\dilatonPole }
$$
\noindent where $g_D=-{M^{3/2}\over
\sqrt{N}}\sqrt{(96\pi)/(1-\lambda^2)}$ is determined from the
infrared behavior of the scalar vertex in Eq.\WfullLadder
$$\eqalignno{
W(p^2,k_3) =V(p^2,k_3) (1+ \lambda^{eff}_4 B_{CS}(k_3))^{-1} \to
({12M^2\over 1-\lambda^2}){1\over k_3^2}=f_Dg_D {1\over k_3^2}
&&\eqnd\gDfrom }
$$
$g_D^2$ also has a pole as $\lambda^2 \to 1$  ~. The boundary
value of $\lambda^2 < 1 $ ~implies a large positive value for
$\lambda_6$ ~for the massive phase as the critical coupling
condition in Eq.\critical ~ would require $\lambda^2=4$ ~without
the $\lambda_6$ ~deformation.

\vskip .3cm
\subsection {\bf ~~~~~~~~~~~Explicit Breaking of Scale Invariance and the
pseudo-dilaton}

\sslbl\ssCd
 \vskip .2cm

The massless dilaton in Section 3.4 was found in the massive phase
of the theory with spontaneously broken conformal invariance.
This phase only exists at the critical coupling~
$\lambda^2+{\lambda_6\over 8\pi^2} = 4$
~in the conformal limit when $\mu_R^2=0$ ~and~ $\lambda_{4R}=0$.
~However, we may also study the subcritical theory if we add
terms which explicitly break the conformal symmetry
but stabilize the vacuum of the massive phase.
Without such symmetry breaking the
subcritical theory has only the massless phase.
If we are close to the critical coupling we can find
solutions where the explicit breaking is small,
the dilaton develops a small mass and the scalar bound state
may be considered as a pseudo-dilaton.

We will consider now the massive phase of the
near critical theory where the explicit breaking terms,~
$\mu_R^2$ ~and~ $\lambda_{4R}$, are chosen so the gap
equation in \SigmaRen ~has a stable solution
with the mass of the $\phi$ ~boson equals $M$.
We will expand solutions in $\lambda_{4R}~,~\mu_R^2$ and in the deviation from
criticality, $\delta \ll 1$~, defined by

$$\eqalignno{ {1\over 4}(\lambda^2 + {\lambda_6\over 8\pi^2}) = 1-\delta
&&\eqnd\explicitBr }$$

\noindent The gap equation~ \SigmaRen ~takes the suggestive form

$$\eqalignno{  M^2\delta = -\lambda_{4R}({M\over 4\pi}) + \mu_R^2 &&\eqnd\GapEqRen }$$

We can now explore the pole in the
scalar current correlator in Eq.\BCSdef   ~where $B_{CS}(k_3)$
~is expanded as in Eq.\BCSexpan  ~and $\lambda_4^{eff}$ ~is given in Eq.\lambdaEFF

$$\eqalignno{ \lambda^{eff}_4=\lambda_{4R}-2\pi M(\lambda^2+{\lambda_6\over 8\pi^2})
 =\lambda_{4R}-8\pi M(1-\delta)
&&\eqnd\lambdaEff}$$

The expansion of the denominator at low
momenta $k^2/M^2 \ll 1$ and $\lambda_{4R}/M ~,~ \delta \ll 1$ becomes

$$\eqalignno{ 1+\lambda_4^{eff}B_{CS}(k) &=
1+\{\lambda_{4R}-8\pi M(1-\delta)\}({1\over 8\pi})\{{1\over M}
-({1\over 12})({k^2\over M^3})(1-\lambda^2)\}+ . . . \cr
&= ({\lambda_{4R}\over 8\pi M}) + \delta
+ ({1\over 12})({k^2\over M^2})(1-\lambda^2) + . . .
&\eqnd\denominator}$$

From Eq.\denominator ~we can read off the mass of the pseudo-dilaton

$$\eqalignno{ M_{pD}^2 = ({12 M^2\over (1-\lambda^2)})
[({\lambda_{4R}\over 8\pi M})+\delta ]
&&\eqnd\PSDilatonMass}$$

For the expansion to make sense, the
pseudo-dilaton mass must be small
compared to the scalar boson mass, $M$.
The two terms in Eq.\PSDilatonMass ~can
be of the same order as the explicit breaking by $\lambda_{4R}$
 can be small only if~ $\delta$ ~is also
 small.

For the near critical theory, the $U(N)$ singlet scalar bound
state acts like a pseudo-dilaton.
However, we also note the presence of a pole at $\lambda^2=1$
in the expressions for the dilaton mass in Eq.\PSDilatonMass
. The limit~  $\lambda \to 1$
~is thought to reflect the boundary of the
Chern-Simons theory \refs \AharonyI \refs\Jain  ~and the
dilatonic interpretation  of the scalar boson bound state is lost.

\vskip .2cm
\noindent It may be interesting to consider the case when
we leave $\mu_R=0$ and take $\lambda_{4R} < 0$.
The gap equation in Eq. \SigmaRen ~has, in addition to the obvious
massless solution, a massive solution with

$$\eqalignno{ {\sqrt |\Sigma|}=M=-{1\over 4\pi}{\lambda_{4\pi}\over \delta}
 &&\eqnd\mass }$$

 \noindent The induced effective coupling in Eq.\lambdaEFF ~is now

 $$\eqalignno{ \lambda^{eff}_4=\lambda_{4R}-2\pi M(\lambda^2+{\lambda_6\over 8\pi^2})
 =-8\pi M(1-\half {\delta})
&&\eqnd\lambdaEFFBr}$$

We note that  Eq.\JJexpanded ~implies now a  pseudo-dilaton solution with
a mass given by

$$\eqalignno{M^2_{pD}= \delta ({6M^2\over 1-\lambda^2})
&&\eqnd\psDilMass}$$

\vskip .3cm
\subsection {\bf ~~~~~~~~~~~The  $<J_0 J_0 J_0>$ Correlator and the Three
Dilatons Interaction}

\sslbl\ssCe
 \vskip .2cm

\noindent Here we calculate the correlation  $<J_0(k) J_0(k')~J_0(-k-k')>$
in the massive ($M\neq 0$) phase and
again taking advantage of an expansion in $k_3/M$.

The first four diagrams (a,b,c,d) in Fig. 5
combine to give for the vertex:
$$\eqalignno{ V^{(a-d)}(k,k',-k-k')
 &=  {N\over 16\pi}{1\over M^3} \{1+{(\lambda^2-{3\over 2})\over 12M^2}(k_3^2+k_3'^2+(k_3+k_3')^2)
+{\cal O}({(k_3,k'_3)^4\over M^4})   \}
\cr &&\eqnd\vertexJJJabcd }$$
\noindent when diagram e is added one obtains in the same limit:
$$\eqalignno{ V_{e}
&= -(\lambda^2+{\lambda_6\over 8\pi^2}) {N\over 64\pi}{1\over M^3} \{1+{(\lambda^2-1)\over 12M^2}
(k_3^2+k_3'^2+(k_3+k_3')^2)
+{\cal O}({(k_3,k'_3)^4\over M^4})   \}
\cr &&\eqnd\vertexJJJe }$$

\midinsert
\epsfxsize=120.mm
\epsfysize=20.mm
\centerline{\epsfbox{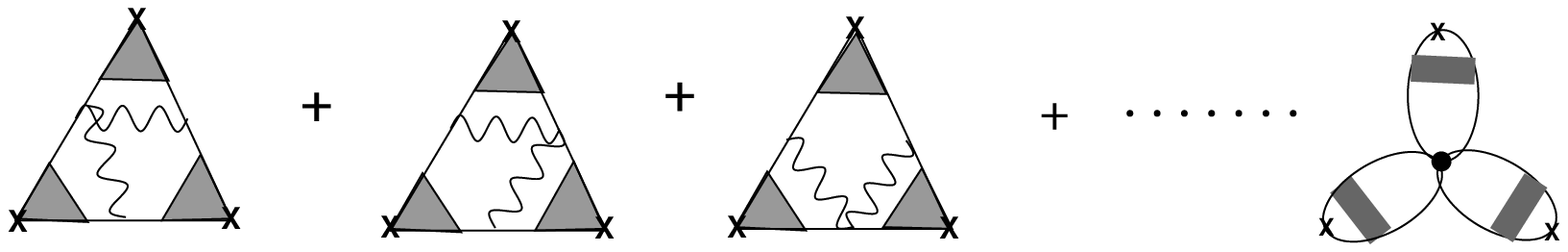}}
\figure{3.mm}{$<J_0(k) J_0(k')~J_0(-k-k')>$}
\endinsert

In view of the gap equation in \gapEq, the expansion in powers of  $k_3/M$ is valid only if
in Eqs.\vertexJJJabcd, \vertexJJJe ~we choose $\lambda^2+{\lambda_6\over 8\pi^2} =4$
and thus $M^2\neq 0$.
At this critical point, the small $k_3/M$~ contribution of order $\lambda^2$
in \vertexJJJabcd ~and~
\vertexJJJe cancels out and we are left with
$$\eqalignno{ V_{{\rm massive ~phase}}
&= - {N\over 16\pi}{1\over 24 M^5}
(k_3^2+k_3'^2+(k_3+k_3')^2)+{\cal O}({(k_3,k'_3)^4\over M^7})
&\eqnd\Vcritical }$$

The
full
three
current
correlation
function
is
obtained
by
adding
the
bubble
iterations
that
depend
on $\lambda_4^{eff}$
to
the
vertex
in
Eq.\Vcritical
$$\eqalignno{ &<J_0(k) J_0(k')~J_0(-k-k')>\cr
&= V(k,k',-k-k') (1+\lambda_4^{eff}B_{CS}(k))^{-1}
(1+\lambda_4^{eff}B_{CS}(k'))^{-1}(1+\lambda_4^{eff}B_{CS}(-k-k'))^{-1}
\cr && \eqnd\JJJcor }$$

Inserting here the above values of $V(k,k',-k-k')$,
$B_{CS}$ and $\lambda_4^{eff}$ in the
massive phase we find a
simple
result
to
leading
order
in
the
momentum
expansion

$$\eqalignno{ &<J_0(k) J_0(k')~J_0(-k-k')>\cr
&= -{N\over 384\pi}{1\over  M^5}
(k_3^2+k_3'^2+(k_3+k_3')^2) ( {k^2\over 12 M^2 }(1-\lambda^2))^{-1}\cr
& \ \ \ \ \ \ \ \ \ \ ( {k'^2\over 12 M^2 }(1-\lambda^2))^{-1}
( {(k+k')^2\over 12 M^2 }(1-\lambda^2))^{-1}\cr
&=-{9N\over 2\pi}{M\over (1-\lambda^2)^3}\{ {1\over k^2 k'^2}+{1\over k^2(k+k')^2}
 +{1\over k'^2(k+k')^2}\} \cr
 & \ \ \ \ \ \ \ \ \ \  +{\cal{O}}({M^3 \over k^2})
& \eqnd\JJJcorfinal }$$


From Eq.\JJJcorfinal ~we can infer the three dilaton coupling. As
the triple pole contribution cancels exactly, the nonderivative
three dilaton coupling constant vanishes. We can interpret
Eq.\JJJcorfinal ~as evidence for a derivative interaction for the
dilaton reflecting its nature as the Goldstone boson of
spontaneously broken conformal symmetry.

\noindent The dilaton self interaction can be now defined in the
effective Lagrangian
$$\eqalignno{ &{\cal L}_3D = g_{3D}
({ \del}_\mu D\cdot{ \del}_\mu D ) D &\eqnd\DDDeffective }$$

\noindent One identifies in Eq.\JJJcorfinal ~the three dilaton coupling
$g_{3D}$ from
$$\eqalignno{ &f_D^3 g_{3D} = -N{9\over 2\pi}
{M\over (1-\lambda^2)^3} &\eqnd\fDgD }$$
\noindent and using Eq.\fD~ $g_{3D}$ is given by
$$\eqalignno{ &g_{3D} = -
{\sqrt {6\pi\over N M (1-\lambda^2)^3}} &\eqnd\gD }$$

\vskip 1cm
\section {\bf ~~~~~~~~~~~~Summary and Conclusions}

\sslbl\ssD

\vskip -.6cm

We presented in this paper several aspects of
spontaneous breaking of scale
invariance in the large N limit of a three dimensional
 Chern-Simons theory coupled to a scalar field in
the fundamental representation of $U(N)$. In the presence of self
interaction term $\lambda_6 (\phi^\dagger\cdot\phi)^3$, a massive
phase is found at a critical value of the combination of
$\lambda_6$ and 't Hooft's  coupling, $\lambda=N/\kappa$ , given
by the condition $\lambda^2+\lambda_6/8\pi^2=4$. At this critical
value, conformal symmetry is spontaneously broken and a massless
dilaton appears in the ground state.

Our explicit results are exact to leading order in the large N expansion and make
use of the light-cone gauge for the Chern-Simons gauge field.
At the critical coupling, we have shown that gap equation for the
scalar self-energy has a massive solution in addition to the
massless solution of the normal conformal phase.  We have, then,
computed the explicit solutions for the full scalar vertex function.
Using these results, we are able to compute the exact scalar current
correlation functions for the two point, $<JoJo>$, and three point,
$<JoJoJo>$ correlators.  These computations are exact in the large
N limit to all orders in 't Hooft coupling, $\lambda=N/\kappa$,
and the $\phi^6$ coupling $\lambda_6$ ~that is,
in turn, fixed by the criticality condition.
From these correlation functions we have inferred the properties of the
composite $U(N)$ ~singlet dilaton that arises in the theory due to the spontaneously broken
scale symmetry in the massive phase.   We have determined the effective dilaton
decay constant, $f_D$ , the coupling of the dilaton to the scalar boson, $g_D$
and the self-coupling of the dilaton, $g_{3D}$.   These couplings are all
singular at $\lambda=1$ which is thought to be the boundary value of
the physical theory.   Finally, we have also explored the near
critical theory and found that there can exist phases with
both spontaneous and explicit scale symmetry breaking.
In these phases, the composite $U(N)$ singlet bound state is
massive and plays the role of a pseudo-dilaton.

Chern-Simons gauge theories in  three dimension have many
remarkable properties which are only
now beginning to be explored with the use of
the large N expansion and other methods.
As in the case of the
massless conformal symmetric phase~\Wadia\AharonyI\Chang, also in the massive,
spontaneously broken conformal invariance phase, the $U(N)$ ~singlet
sector  is conjectured to be dual to a parity
broken version of Vasiliev's higher
spin theory  on $AdS_4$ in the bulk.
It would be interesting to explore the implications for the bulk
four dimensional AdS dual description of the massive phase. This
is an open problem whose solution is not known at this point. In
particular it is unknown whether the bulk theory is just a
modification of Vassiliev's theory or whether new fields are
required. The role of the AdS dual of the composite dilaton
behavior and meaning of the critical point for the specific
combination of coupling constants would be interesting to explore.
This could have implications for AdS duals of near conformal
theories in four dimensions and, perhaps, the approximate AdS
duals used to describe QCD in the large N limit.  These
interesting problems are beyond the scope of the present
investigation.

\vskip 1.5cm

\centerline{\bf Acknowledgements }

\noindent MM thanks Fermilab and Saclay theory groups for their
warm hospitality while different parts of this work were done.
This work was completed while WB and MM visited CERN and we thank
the CERN TH group for their warm hospitality. We would like to
thank Ofer Aharony, Eliezer Rabinovici and Jean Zinn-Justin for
useful discussions and correspondence on some of the issues
discussed here. Fermilab is operated by Fermi Research Alliance,
LLC under Contract No. DE-AC02-07CH11359 with the United States
Department of Energy.

\vfill\eject

\listrefs

\vfill\eject

\bye